\documentclass[12pt]{article}
\usepackage{latexsym,epsfig,graphicx}

\def\o{\omega}

\def\S{\Sigma}

\def\be{\begin{equation}}
 \def\ee{\end{equation}}
 \def\bea{\begin{eqnarray}}
 \def\eea{\end{eqnarray}}

 \def\o{\omega}
 
\newcommand{\fr}{\frac}
\newcommand{\pr}{\prime}
\newcommand{\pp}{{\prime \prime}}

\def\2{\frac{1}{2}}
\def\4{\frac{1}{4}}

\catcode`\@=11

\@addtoreset{equation}{section}

\def\@normalsize{\@setsize\normalsize{15pt}\xiipt\@xiipt
\abovedisplayskip 14pt plus3pt minus3pt%
\belowdisplayskip \abovedisplayskip
\abovedisplayshortskip  \z@ plus3pt%
\belowdisplayshortskip  7pt plus3.5pt minus0pt}
\def\small{\@setsize\small{13.6pt}\xipt\@xipt
\abovedisplayskip 13pt plus3pt minus3pt%
\belowdisplayskip \abovedisplayskip
\abovedisplayshortskip  \z@ plus3pt%
\belowdisplayshortskip  7pt plus3.5pt minus0pt
\def\@listi{\parsep 4.5pt plus 2pt minus 1pt
            \itemsep \parsep
            \topsep 9pt plus 3pt minus 3pt}}

\def\underline#1{\relax\ifmmode\@@underline#1\else
        $\@@underline{\hbox{#1}}$\relax\fi}
\@twosidetrue \relax

\catcode`@=12

\catcode`\@=11

\def\section{\@startsection{section}{1}{\z@}{3.5ex plus 1ex minus
   .2ex}{2.3ex plus .2ex}{\large\bf}}

\def\ps@headings{\def\@oddfoot{}\def\@evenfoot{}
\def\@oddhead{\hbox{}\hfill
        \makebox[.5\textwidth]{\raggedright\ignorespaces --\thepage{}--
        \hfill }}
\def\@evenhead{\@oddhead}
\def\subsectionmark##1{\markboth{##1}{}}
}

\ps@headings

\catcode`\@=12

\begin{document}

\begin{titlepage}
\begin{flushright}
January 2008
\end{flushright}

\vspace{0.30in}
\begin{centering}

{\large {\bf Phase Transitions in Charged Topological-AdS  Black
Holes}}

\vspace{0.7in}
 {\bf George Koutsoumbas$^{*}$, \,\,\,\,
Eleftherios Papantonopoulos $^{\flat}$ }\,\,\,\,\\
\vspace{0.2in}

{\em Department of Physics, National Technical University of Athens,\\
Zografou Campus GR 157 73, Athens, Greece}\\
 \vspace{0.14in}
 and \\
  \vspace{0.2in}
{\bf George Siopsis}
 $^{\natural}$ \\
 \vspace{0.2in}
{\em Department of Physics and Astronomy, The University of
Tennessee,\\
Knoxville, TN 37996 - 1200, USA}

\vspace{0.44in}

\vspace{0.04in}

\end{centering}

\vspace{0.6in}

\begin{abstract}

We study the perturbative behaviour of charged topological-AdS
black holes. We calculate both analytically and numerically the
quasi-normal modes of the  electromagnetic and gravitational
perturbations. Keeping the charge-to-mass ratio constant, we show
that there is a second-order phase transition at a critical
temperature at which the mass of the black hole vanishes. We pay
special attention to the purely dissipative modes appearing in the
spectrum as they behave singularly at the critical point.

 \end{abstract}

\begin{flushleft}

\vspace{0.990in}

 $^{*}$~kutsubas@central.ntua.gr \\
$^{\flat}$~lpapa@central.ntua.gr\\
$^\natural$~siopsis@tennessee.edu

\end{flushleft}
\end{titlepage}

\section{Introduction}

It is well known that if a black hole is initially perturbed, the
surrounding geometry will start vibrating into quasi-normal
oscillation modes, whose frequencies and decay times depend only
on the intrinsic features of the black hole itself, being
insensitive to the details of the initial perturbation. The
radiation associated with these modes is expected to be seen with
gravitational wave detectors in the coming years, giving valuable
information on the properties of black holes. For these reasons,
quasi-normal modes (QNMs) of black holes in asymptotically flat
spacetimes have been extensively studied (for reviews,
see~\cite{KS,N}).

The Anti-de Sitter - conformal field theory (AdS/CFT)
correspondence has led to an intensive investigation of black hole
QNMs in asymptotically AdS spacetimes. Quasi-normal modes in AdS
spacetime were first computed for a conformally invariant scalar
field, whose asymptotic behaviour is similar to flat
spacetime~\cite{CM}.  Subsequently, motivated by the AdS/CFT
correspondence, Horowitz and Hubeny made a systematic computation
of QNMs for scalar perturbations of Schwarzschild-AdS (S-AdS)
spacetimes~\cite{HH}.  Their work was extended to electromagnetic
and gravitational perturbations of S-AdS black holes in~\cite{CL}.
The study of scalar perturbations was further extended to the case
of Reissner-Nordstr\"om-AdS (RN-AdS) black holes in~\cite{WLA}.
Finally, the QNMs of  scalar, electromagnetic and gravitational
perturbations of RN-AdS black holes were presented
in~\cite{kokkotas} using the results of~\cite{MM}.

The QNMs of AdS black holes have an interpretation in terms of the
dual conformal field theory (CFT) \cite{cft}.
 According to the AdS/CFT
correspondence, a large static black hole in AdS corresponds to an
 (approximately)
thermal state in the CFT. Perturbing the black hole corresponds to
perturbing this thermal state, and the decay of the perturbation
describes the return to thermal equilibrium. So we obtain a
prediction for the thermalization timescale in the strongly
coupled CFT. In ref.~\cite{HH} it was shown that the QNMs for the
scalar perturbations of large Schwarzschild-AdS black holes scaled
with the temperature and it was argued that the perturbed system
in the dual description will approach to thermal equilibrium of
the boundary conformal field theory. However, when the black hole
size is comparable to the AdS length scale there is a clear
departure from this behaviour. It was then conjectured that this
behaviour may be connected with a Hawking-Page phase transition
\cite{phase1, phase2} which occurs when the temperature lowers.

These results were further confirmed in
\cite{CL,WLA,kokkotas,Konoplya}. However, the behaviour of QNMs
for small black holes is still poorly understood. Another
interesting finding of the electromagnetic and gravitational
perturbations is that  purely dissipative modes appear in the
spectrum which are pure imaginary QNMs. In such perturbed
classical backgrounds the presence of these dissipative modes
indicate that the boundary theory reaches thermal equilibrium with
no oscillations. It was shown in \cite{CL} that for axial
perturbations of Schwarzschild-AdS black holes these highly damped
modes scale as the inverse of the black hole radius and this
behaviour persisted in the case of Reissner-Nordstr\"{o}m-AdS
black holes \cite{kokkotas}.

It was observed in \cite{Martinez:2004nb} that  for small black
holes as we lower the temperature to a critical value there is a
phase transition of a vacuum topological black hole towards a
hairy black hole (MTZ). This claim was supported in
\cite{Koutsoumbas:2006xj} by calculating the QNMs of
electromagnetic perturbations of the MTZ black hole and
topological black holes. It was found that there is a change in
the slope of the QNMs as we decrease the value of the horizon
radius below a critical value, and this change has been attributed
to the phase transition.  It was also observed in
\cite{Koutsoumbas:2006xj} that for small
 black holes  the quasi-normal frequencies
converge toward the imaginary axis, i.e., their real part
decreases and after the first few quasi-normal frequencies, it
vanishes, indicating that for black hole radius smaller than the
AdS length scale  there are only a finite number of QNMs. It was
shown that the finite number of such modes for small horizons
 is due to the existence of bound states behind the
horizon, which is an unobservable region. Further evidence that
the behaviour of QNMs may provide indications of a phase
transition was provided in \cite{Shen:2007xk}.

Topological black holes~\cite{mann}-\cite{Birmingham:2006zx},
having hyperbolic horizons, introduce new features not present in
spherical black holes. In~\cite{Emparan:1999gf} it was shown that
hyperbolic black holes can be described as thermal Rindler states
of the dual conformal field theory in flat space. It was also
found that, for small topological black holes of the size of the
AdS length scale, the entropy at strong coupling is larger than
the entropy obtained from field theory at lowest perturbation
order at weak coupling. One possible explanation put forward was
that there is a phase transition of topological black hole to
vacuum AdS space at a critical temperature, which however was not
observed in~\cite{Emparan:1999gf} (see also~\cite{Birmingham,
Emparan:1998pf}).

In this work we make a detailed study of electromagnetic and
gravitational perturbations of charged topological black holes
(CTBH) in AdS space. Studying these perturbations of the
background geometry, we show that the second-order phase
transition observed at a critical temperature
in~\cite{Koutsoumbas:2006xj} occurs in more general configurations
including charge. We calculate both analytically and numerically
the QNMs of axial and in some cases the polar perturbations of
large and small black holes. We find that for large black holes
the QNMs irrespectively of the value of the charge, exhibit a
negative slope. However, if the value of the black hole radius is
smaller than the AdS length scale and the charge is small, the
propagating QNMs (whose frequencies have a non-vanishing real
part) of both axial and polar perturbations are finite with a
positive slope. We attribute this behaviour to a second order
phase transition which occurs as the temperature approaches a
critical value (or the black hole radius is approaching the length
scale of the AdS space). We also find that, in the case of small
black holes, if we increase the charge, the number of propagating
QNMs is again infinite and a part of positive slope coexists with
the negative slope frequencies. This observation indicates that in
the case of small black holes, the charge plays the r$\hat{o}$le
of an order parameter and according to the AdS/CFT correspondence
we expect that the dual boundary theory is described by a thermal
state with coexisting phases.

 We  also study the purely dissipative modes appearing in the spectrum, both
 analytically and numerically of electromagnetic and gravitational perturbations of
 the charged topological black holes. For large black holes the
 purely dissipative modes of both axial and polar perturbations scale
 linearly with temperature. Also the intermediate black holes to a
 good approximation depend linearly on the temperature.
 This is the expected behaviour for large and intermediate black
 holes and they agree with the normal QNMs behaviour discussed in~\cite{HH}.
If the black hole radius is smaller than the AdS length scale then
we find a clear departure from linearity with temperature. If the
charge is small the purely dissipative modes scale with the
temperature according to $a+b/(T-T_0)$ where $a, b$ are constants.
Then, for a fixed charge to mass ratio we observed an infinite
change of slope at $T=T_{0}$ signaling a second order phase
transition. As we increase the charge the temperature dependence
changes drastically.

The paper is organized as follows. In section \ref{sec2} we review
the basic properties of the topological black holes and we discuss
their thermodynamics. In section \ref{sec3} we present the
analytical calculations of the QNMs of electromagnetic and
gravitational perturbations of the CTBH. In section \ref{sec4} we
study numerically their behaviour and in section \ref{sec5} we
discuss in detail the purely dissipative modes appearing in the
spectrum. Finally, section \ref{sec6} contains our summary.

\section{Thermodynamics}
\label{sec2} We consider the action $$ I=\int
d^{4}x\sqrt{-g}\Big{[} \frac{R+6l^{-2}}{16\pi G}\Big{]}~,$$ where
$l$ is the AdS radius. The presence of
 a negative cosmological constant $(\Lambda=-3l^{-2})$ allows the existence of black holes with a
topology ${R}^{2} \times \Sigma$, where $\Sigma $ is a
two-dimensional manifold of constant curvature. These black holes
are known as topological black holes
\cite{mann}-\cite{Birmingham}. The simplest solution of this kind,
when $\Sigma $ has negative constant curvature, reads
\begin{equation}
ds^{2}=-\left( -1+\frac{r^{2}}{l^{2}}-\frac{2G\mu }{r}\right)
dt^{2}+\frac{ dr^{2}}{\left( -1+\frac{r^{2}}{l^{2}}-\frac{2G\mu
}{r}\right) }+r^{2}d\sigma ^{2}\;,  \label{Top-BH-Einstein}
\end{equation}
where $%
d\sigma ^{2}$ is the line element of $\Sigma $, which is locally
isomorphic to the hyperbolic manifold $H^{2}$ and $\Sigma $ must
be of the form
\[
\Sigma =H^{2}/\Gamma {\quad }\mathrm{{with\quad }\Gamma \subset
O(2,1)\;,}
\]
where $\Gamma $ is a freely acting discrete subgroup (i.e.,
without fixed points).

The configurations (\ref{Top-BH-Einstein}) are asymptotically
locally AdS spacetimes. It has been  shown in
\cite{Gibbons:2002pq} that the massless configurations where
$\Sigma $ has negative constant curvature are stable under
gravitational perturbations. More recently the stability of the
topological black holes was discussed in \cite{Birmingham:2007yv}.

If we introduce to the above action an electromagnetic field
$$-\frac{1}{16\pi}\int d^{4}x\sqrt{-g}F^{\mu \nu}F_{\mu \nu} $$
then the metric of a charged topological black hole is given by
  \be
 ds^{2}=-f(r)dt^{2}+\frac{dr^{2}}{f(r)}+r^{2}d\sigma^{2}~,\,\,
 \,\,\,\, f(r)=r^{2}-1-\frac{2G\mu}{r}+\frac{q^{2}}{r^{2}}~, \ee where we have fixed the length of the AdS space to
 $l=1$.
 The horizon $r_{+}$ is specified from \be q^{2}=2G\mu
 r_{+}+r^{2}_{+}-r^{4}_{+}~.\label{horizon}\ee
  Define the charge to mass
 ratio by \be \frac{q^{2}}{(G\mu)^{2}}=\lambda^{2}~.\label{ratio}\ee Then
 using (\ref{horizon}) we have\footnote{There is another possibility,
\[ G\mu=\frac{r_{+}(r^{2}_{+}-1)}{1-\sqrt{1-\lambda^{2}
(r^{2}_{+}-1)}} \] We shall not discuss it here because it
corresponds to configurations which do not approach the critical
point $r_+=1$.} \be
G\mu=\frac{r_{+}(r^{2}_{+}-1)}{1+\sqrt{1-\lambda^{2}
(r^{2}_{+}-1)}}~. \label{rathorizon} \ee We consider an electric
potential at the horizon $\Phi=-q/r_{+}$ and the electric charge
$Q=\sigma q/4\pi$. The temperature, the entropy and the mass of
the CTBH are given by \bea T&=& \fr{f^\pr(r_+)}{4 \pi} =
\frac{3r^{4}_{+}-r^{2}_{+}-q^{2}}{4 \pi r^{3}_{+}}~, \nonumber
\\S&=&\frac{\sigma r^{2}_{+}}{4 G}~ ,\nonumber \\
M&=&\frac{\sigma \mu}{4
\pi}=\frac{\sigma(r^{4}_{+}-r^{2}_{+}+q^{2})}{8 \pi G
r_{+}}~.\label{relations} \eea With the help of $$dM =
\fr{\sigma(3 r_+^2-1-\fr{G q^2}{r_+^2})}{8 \pi G} dr_+ +\fr{\sigma
q}{4 \pi r_+} dq$$ it is easy to verify that the law of
thermodynamics \be dM=TdS-\Phi dQ \label{dm}\ee is valid. The free
energy is \be F=M-TS+Q\Phi=-\frac{\sigma(r^{3}_{+}+r_{+})}{16\pi
G}-\frac{\sigma q^{2}}{16\pi r_+}=-\frac{\sigma}{16\pi
G}\left[r_+^3+r_++\frac{\lambda^2 (G \mu)^{2}}{r_+} \right],\ee
where $G \mu$ should be replaced by its value in eq.
(\ref{rathorizon}) and then \be dF=-SdT+Qd\Phi \label{df} \ee is
easily verified. Notice now that the critical temperature \be T_0=
\frac{1}{2\pi} \ee corresponds to $r_+=1.$  We have already given
the functions $F(r_+)$ and $T(r_+),$ so we can readily Taylor
expand $F(T)$ about $T_0$ and express the necessary derivatives
through the formulae $$\fr{d F}{d T} =\fr{F^\pr}{T^\pr}, \fr{d^2
F}{d T^2} = \fr{T^\pr F^\pp-F^\pr T^\pp}{T^{\pr 3}}, \fr{d^3 F}{d
T^3}  = \fr{3 F^\pr T^{\pp 2} + T^{\pr 2} F^{(3)} -3 T^\pr T^\pp
F^\pp-F^\pr T^\pr T^{(3)}}{T^{\pr 5}}, \dots,$$ where primes
denote differentiations with respect to $r_+.$ The expansion of
the free energy finally reads: \bea F&=&-\frac{\sigma }{8 \pi
G}\Big{(}1+2(T-T_{0})\pi
+(2+\lambda^{2})(T-T_{0})^{2}\pi^{2}\nonumber
\\ &+&\frac{(2+3\lambda^{2})(1+\lambda^{2})}{2}
(T-T_{0})^{3}\pi^{3}+...\Big{)}~.\label{energyctbh}\eea It is
interesting to consider how formulae (\ref{horizon}) -
(\ref{energyctbh}) simplify in two important special cases, namely
when $\lambda^2$ takes on the values $-1$ or $0.$ Notice that
having chosen a definite value for $\lambda$ means that $q$ and
$r_+$ are no longer independent.

  ${\bf (a)~~~\lambda^2=-1}$~~(CTBH with
scalar hair~\cite{Martinez:2005di})

\be ds^{2}=-f(r)dt^{2}+\frac{dr^{2}}{f(r)}+r^{2}d\sigma^{2}~,\,\,
\,\,\,\, f(r)=r^{2}-\left(1+\frac{G\mu}{r}\right)^2~. \ee The
horizon $r_{+}$ is specified from \be G\mu =
r_+(r_+-1)~,\label{MShorizon}\ee and the relevant quantities
become \be Q=\frac{\sigma q}{4\pi}\ee \bea T&=& \fr{f^\pr(r_+)}{4
\pi} = \frac{2 r_{+}-1}{2 \pi}~, \nonumber
\\S&=&\frac{\sigma r^{2}_{+}}{4 G}~ \Rightarrow
dS=\frac{\sigma r_+ dr_+}{2 G},\nonumber \\ M&=&\frac{\sigma
\mu}{4 \pi}=\frac{\sigma(r^{2}_{+}-r_{+})}{4 \pi G}.
\label{MSrelations} \eea \be dM=TdS-\Phi dQ~, \label{MSdm}\ee \be
F=M-TS+Q\Phi=-\frac{\sigma r^{2}_{+}}{8\pi G}~,\ee \be
dF=-SdT+Qd\Phi~, \label{msdf} \ee \bea F&=& -\frac{\sigma}{8 \pi
G} \Big{(}1+2(T-T_{0})\pi
+(T-T_{0})^{2}\pi^{2}+O(T-T_{0})^{4}\nonumber
\Big{)}~.\label{energyMS}\eea

${\bf (b)~~~\lambda^2=0}$~~(Uncharged topological black hole) with
\be f(r)=r^{2}-1-\frac{2G\mu}{r}~, \ee \be 2G\mu=
r^{3}_{+}-r_{+}~,\label{TBHhorizon}\ee \be Q=\sigma q/4\pi=0~.\ee
\bea T&=& \fr{f^\pr(r_+)}{4 \pi} = \frac{3r^{2}_{+}-1}{4 \pi
r_{+}}~, \nonumber
\\S&=&\frac{\sigma r^{2}_{+}}{4 G}~
\Rightarrow dS=\frac{\sigma r_+ dr_+}{2 G}~,\nonumber \\
M&=&\frac{\sigma \mu}{4 \pi}=\frac{\sigma(r^{3}_{+}-r_{+})}{8 \pi
G}\Rightarrow dM = \fr{\sigma(3 r_+^2-1)}{8 \pi G} dr_+
~.\label{TBHrelations} \eea \be dM=TdS~, \label{TBHdm}\ee \be
F=M-TS=-\frac{\sigma(r^{3}_{+}+r_{+})}{16\pi G}~,\ee \be
dF=-SdT\label{TBHdf}~, \ee \bea F&=&-\frac{\sigma }{8 \pi
G}\Big{(}1+2(T-T_{0})\pi +2(T-T_{0})^{2}\pi^{2} +
(T-T_{0})^{3}\pi^{3}+...\Big{)}~.\label{energycTBH}\eea

As an application of the above considerations if we assume that
\be f(r)= r^{2} %
-\left( 1+\frac{G\mu }{r}\right) ^{2}\ee solves a four dimensional
gravitational action describing a gravitational field coupled to a
scalar and to an EM field (charged MTZ black hole
~\cite{Martinez:2005di}), calculating the quantity \be
F_{CTBH}-F_{CMTZ}=-\frac{\pi ^{2} \sigma
(2+3\lambda^{2})(1+\lambda^{2})}{16G } (T-T_{0})^{3}\pi^{3}+...\ee
we see that it is changing sign as we cross the critical
temperature, hence we have a phase transition of the charged MTZ
black hole to CTBH.

\section{Gravitational and Electromagnetic Modes}
\label{sec3} The radial wave equation for gravitational
perturbations in the black-hole background can be cast into a
Schr\"odinger-like form \cite{Kodama}, \be\label{sch}
  -\frac{d^2\Psi}{dr_*^2}+V[r(r_*)]\Psi =\o^2\Psi \;,
\ee in terms of the tortoise coordinate defined by
\be\label{tortoise}
  \frac{dr_*}{dr} = \frac{1}{f(r)}\;.
\ee The potential $V$ is determined by the type of perturbation.
For axial perturbations, we have
\be\label{eqVV} V(r) = V_{\mathsf{A}}^{\pm} (r) \equiv f(r)
\left\{ \frac{\Lambda}{r^2} - \frac{3G\mu [ 1\pm \Delta ]}{r^{3}}
+ \frac{4 q^2}{r^{4}} \right\}~, \ee where \be \Delta =
\sqrt{1+\frac{4}{9} (\Lambda+2) \lambda^2 } \ \ , \ee and
$\Lambda$ are the eigenvalues resulting from the the solution of
Laplace-Beltrami equation $\Delta \psi=-\Lambda \psi$ on a space
which has constant negative curvature and which is of the form
$P_{-\frac{1}{2}\pm i \xi}^m (\cosh r) e^{i m \phi}(\cosh r)$ and
the corresponding eigenvalues read: $\Lambda = -l (l+1)$ with
$l=-\frac{1}{2}\pm i \xi,$ so that $\Lambda = \frac{1}{4}+\xi^2.$

For polar perturbations we have, \be\label{eqPS} V(r) =
V_{\mathsf{P}}^\pm (r) \equiv \frac{f(r) U_\pm (r)}{r^2 (\Lambda +
2 + 3(1\pm\Delta) \frac{G\mu}{r} )^2}~, \ee where \bea U_\pm (r)
&=& 18 (1\pm\Delta)^2 G^2\mu^2 + \frac{81}{2}
\frac{1-\Delta^2}{\Lambda+2}
 (1\pm\Delta)^2 \frac{G^4\mu^4}{r^4} + 9(1\pm\Delta)^2 (5\mp 3\Delta) \frac{G^3\mu^3}{r^3} \nonumber\\
&& + 18 (1\pm\Delta) (\Lambda +2) \frac{G^2\mu^2}{r^2} +
3(\Lambda+2)^2 (1\pm\Delta) \frac{G\mu}{r} + \Lambda
(\Lambda+2)^2~. \eea
We shall refer to the modes corresponding to the potential $V^-$
(axial or polar) as $Z_1$ modes whereas those corresponding to
$V^+$ will be referred to as $Z_2$ modes.

The potential for polar perturbations (\ref{eqPS}) has a
singularity at \be\label{eqsing} r = r_0 = -
\frac{3(1\pm\Delta)G\mu}{\Lambda+2} \ee in addition to the
standard singularities $r=r_+,\infty$. Remarkably, at the
singularity (\ref{eqsing}), \be U_\pm (r_0) = 2(\Lambda+2)^2
f(r_0)~, \ee resulting in the behaviour of the wavefunction \be
\Psi \sim (r-r_0)^\alpha \ \ , \ \ \ \ \alpha = 2,-1~. \ee It
turns out that at the boundary ($r\to\infty$), the wavefunction
has the same behavior with $\alpha = -1$. Precisely, \be \Psi \sim
(1-r_0/r)^{-1} \ee so it does not obey the Dirichlet boundary
condition $\Psi = 0$. Instead, it obeys the {\em Robin} boundary
condition \cite{bibRobin,bibsio} \be r^2 \Psi' + r_0 \Psi =0 \ \ \
\ (r\to\infty)~. \ee This is true even in the limit $q=0$ for the
$Z_2$ mode. For the $Z_1$ mode, $r_0=0$ if $q=0$, so we obtain the
standard $\Psi\to 0$ boundary condition. The latter is the
electromagnetic mode. Indeed, if $q=0$, $\Delta = 1$, so \be
U_-(r) = (\Lambda+2)^2 \Lambda~, \ \ \ \
 V_-(r) = \Lambda \frac{f(r)}{r^2}~. \ee
Evidently, the potential vanishes at the horizon ($V(r_+) = 0$,
since $f(r_+)=0$). This is the case for all types of perturbation.


To obtain analytic expressions for the quasi-normal frequencies,
it is convenient to introduce the coordinate \cite{bibsio}
\be\label{eqru} u = \frac{r_+}{r}~. \ee The wave equation
(\ref{sch}) becomes \be\label{eq13} (u^2 \hat f(u) \Psi')' +
\left[ \frac{\hat\omega^2}{u^2 \hat f(u)} - \frac{\hat V}{u^2\hat
f} \right] \Psi = 0 \ \ , \ \ \ \ \hat\omega =
\frac{\omega}{r_+}~,\ee where prime denotes differentiation with
respect to $u$ and we have defined \be\label{eq14} \hat f(u)
\equiv \frac{f(r)}{r_+^2} = \frac{1}{u^2} - \frac{1}{r_+^2} -
\frac{2G\mu}{r_+^{3}} u + \frac{q^2}{r_+^{4}} u^{2} \ \ , \ \ \ \
\hat V(u) \equiv \frac{V(r)}{r_+}~. \ee For the various
potentials, we obtain \be\label{eq15} \hat V_{\mathsf{A}}^\pm (u)
= \hat f(u) \left\{ \hat\Lambda u^2 -
\frac{3G\mu[1\pm\Delta]}{r_+^{3}} u^{3} + \frac{4 q^2}{r_+^{4}}
u^{4} \right\} \ \ , \ \ \ \ \hat\Lambda = \frac{\Lambda}{r_+^2}
\ee and \be\label{eq15p} \hat V_{\mathsf{P}}^{\pm} (u) =
\frac{u^2\hat f(u)\hat U_\pm (u)}{(\tilde\Lambda + 3(1\pm\Delta)
\frac{G\mu}{r_+^3} u)^2}~,\ee
\bea \hat U_\pm (u) &\equiv& \frac{U_\pm (r)}{r_+^6} \nonumber\\
&=& 18(1\pm\Delta)^2 \frac{G^2\mu^2}{r_+^6}
 + \frac{81}{2} \, \frac{1-\Delta^2}{\tilde\Lambda} (1\pm\Delta)^2 \frac{G^4\mu^4}{r_+^{12}}
  u^4 + 9(1\pm\Delta)^2 (5\mp 3\Delta) \frac{G^3\mu^3}{r_+^9} u^3 \nonumber\\
&& + 18 (1\pm\Delta) \tilde\Lambda \frac{G^2\mu^2}{r_+^6} u^2 +
3\tilde\Lambda^2
 (1\pm\Delta) \frac{G\mu}{r_+^3} u
+ \tilde\Lambda^3 - 2\tilde\Lambda^2\frac{1}{r_+^2} ~, \eea where
\be \tilde\Lambda = \hat\Lambda + \frac{2}{r_+^2}~. \ee

\subsection{QNMs of Large Black Holes}

To study the form of quasi-normal modes for large horizons, it is
convenient to factor out the behaviour of the wavefunction at the
horizon ($u=1$),
\be \Psi = (1-u)^{-i \frac{\omega}{4\pi T_H}} F(u)~. \ee The wave
equation becomes
\be\label{sch2} \mathcal{A}_{r_+} F'' +
\mathcal{B}_{r_+,\hat\omega} F' + \mathcal{C}_{\hat\omega ,
\hat\Lambda} F = 0~, \ee where
\bea \mathcal{A}_{r_+} &=& u^2\hat f~, \nonumber\\
\mathcal{B}_{r_+,\hat\omega} &=& (u^2\hat f)' +
2 \frac{i\omega}{4\pi T_H} \frac{u^2 \hat f}{1-u}~, \nonumber\\
\mathcal{C}_{\hat\omega , \hat\Lambda} &=&
\frac{\hat\omega^2}{u^2\hat f} - \frac{\hat V}{u^2\hat f} -
\frac{\omega^2}{(4\pi T_H)^2}\frac{u^2 \hat f}{(1-u)^2} +
\frac{i\omega}{4\pi T_H} \frac{(u^2\hat f)' }{1-u} +
\frac{i\omega}{4\pi T_H}\frac{u^2 \hat f}{(1-u)^2}~.\eea
For $V = V_{\mathsf{A}}^+$ (axial $Z_2$ modes), this wave equation
may be solved for small $\hat\omega$, $\hat\Lambda$. To employ
perturbation theory, write eq.~(\ref{sch2}) as \be (\mathcal{H}_0
+ \mathcal{H}_1) F = 0~, \ee where \bea\label{eqH0} \mathcal{H}_0
F &\equiv& \mathcal{A}_{\infty} F'' + \mathcal{B}_{\infty,0} F' +
\mathcal{C}_{0,0} F~,
\nonumber\\
\mathcal{H}_1 F &\equiv& (\mathcal{A}_{r_+} - \mathcal{A}_\infty )
F'' + (\mathcal{B}_{r_+,\hat\omega} - \mathcal{B}_{\infty,0}) F' +
(\mathcal{C}_{\hat\omega , \hat L} - \mathcal{C}_{0 , 0}) F~. \eea
The zeroth order equation $\mathcal{H}_0 F_0 = 0$ is obtained by
letting $\hat\omega \ , \ \hat\Lambda \to 0$, $r_+\to\infty$ while
keeping $G\mu/r_+^3$ and $q^2/r_+^4$ fixed. We have $\Delta \to
1$, so \be (u^2 \hat f_0(u) F_0')' - \left[ -
\frac{6G\mu}{r_+^{3}} u + \frac{4 q^2}{r_+^{4}} u^{2} \right] F_0
= 0~. \ee Despite its apparent complexity, the acceptable solution
takes a remarkably simple form, \be F_0(u) = u~. \ee The
first-order constraint reads \be\label{eq29a} \int_0^1 F_0
\mathcal{H}_1 F_0 = 0~, \ee which imposes a constraint on the
parameters (dispersion relation) of the form \be\label{eqdispa}
\mathbf{a}_0 -i \mathbf{a}_1 \hat\omega - \mathbf{a}_2
\hat\omega^2 = 0~. \ee After some algebra, we arrive at explicit
expressions for the coefficients, \bea \mathbf{a}_0 &=& -\int_0^1
u^{2} \left[ -\hat\Lambda -\frac{2}{r_+^2} + \frac{3G\mu (\Delta
-1)}{r_+^{3}} u \right]
= \frac{\hat\Lambda + \frac{2}{r_+^2}}{3} + \frac{3G\mu (1-\Delta )}{4}~, \nonumber\\
\mathbf{a}_1 &=& - \frac{r_+}{4\pi T_H} \int_0^1 u^{2} \left[
\frac{2u\hat f_0}{1-u} + \left( \frac{u^2\hat f_0}{1-u} \right)'
\right] = - \frac{r_+}{4\pi T_H} \left. \left( \frac{u^4\hat
f_0}{1-u} \right) \right|_0^1  = 1~. \eea The other coefficient,
$\mathbf{a}_2$ is not needed for the lowest mode. Therefore, \be
i\hat\omega = \mathbf{a}_0 = \frac{\hat\Lambda +
\frac{2}{r_+^2}}{3} + \frac{3G\mu (1-\Delta )}{4r_+^{3}}~. \ee
Explicitly, for large $r_+$, \be\label{eqZ2ana} \omega = -i\,
\frac{(\Lambda + 2)r_+^2 }{6G\mu}~, \ee which is a purely
dissipative mode. Notice that $\omega \propto 1/r_+$, because
$G\mu \propto r_+^3$ for large $r_+$.

For $Z_1$ axial perturbations ($V = V_{\mathsf{A}}^-$), there is
no solution in the small $\hat\omega$ limit. This indicates that
the lowest lying modes are proportional to $r_+$ (so that
$\hat\omega = \omega/r_+$ remains finite as $r_+\to\infty$).
Explicit analytic expressions cannot be obtained in general.
However, as we discuss later, in the case of $q=0$ the
wavefunction may be written in terms of a Heun function leading to
semi-analytic expressions for the frequencies.

The calculation of polar modes is considerably more involved due
to the additional
 singularity of the potential \cite{bibsio}. The spectrum is similar to
  the spectrum of axial modes as evidenced by our numerical calculations (section~\ref{sec4}).
We shall not perform the analytical calculation of these modes for
large black holes as our main focus is on the critical point
($r_+=1$) to which we turn next.

\subsection{QNMs at the Critical Point}

At the critical point ($r_+=1$, $q=0$, $\mu = 0$), the wave
equation for all types of perturbations reduces to \be
((1-u^2)\Psi')' + \left[ \frac{\omega^2}{1-u^2} - \Lambda \right]
\Psi = 0~, \ee whose solutions can be written in terms of
associated Legendre functions. The solution which is well-behaved
at the horizon is \be \Psi(u) = P_{i\xi - 1/2}^{i\omega} (u)~. \ee
To see that it is regular at $u=1$, express it in terms of a
hypergeometric function, \be \Psi(u) = \frac{1}{\Gamma
(1-i\omega)}\ \left( \frac{1+u}{1-u} \right)^{i\omega/2} F(-i\xi +
\frac{1}{2}, i\xi + \frac{1}{2} ; 1-i\omega ; \frac{1-u}{2} )~.
\ee At the boundary, $u\to 0$, it approaches a constant, \be
\Psi(0) = P_{i\xi - 1/2}^{i\omega} (0) = \frac{2^{i\omega}
\sqrt\pi}{\Gamma(\frac{3}{4} - \frac{1}{2} i\xi - \frac{1}{2}
i\omega)\Gamma(\frac{3}{4} + \frac{1}{2} i\xi - \frac{1}{2}
i\omega) }~. \ee Demanding that it vanish, we deduce the
quasi-normal frequencies \be\label{eqana1} \omega_n = \pm \xi - i
\left( 2n + \frac{3}{2} \right) \ \ , \ \ \ \ n = 0,1, 2, \dots
\ee which have finite real part (except in the special case $\xi =
0$).

For explicit expressions, use
\[ F(\alpha,\beta;\gamma; z) = (1-z)^{\gamma -\alpha-\beta} F(\gamma -\alpha,\gamma-\beta; \gamma; z) \]
to write \be \Psi_n (u) = A_n (1-u^2)^{-i\omega_n/2}
F(-i\omega_n+i\xi + \frac{1}{2}, -i\omega_n -i\xi + \frac{1}{2} ;
1-i\omega_n ; \frac{1-u}{2} )~. \ee These hypergeometric functions
are polynomials. Explicitly, \bea \Psi_0 (u)
&=& A_0 (1-u^2)^{-i\xi/2-3/4} u~, \nonumber\\
\Psi_1(u) &=& A_1 (1-u^2)^{-i\xi/2-7/4} u \left[ 1 +
\frac{2+2i\xi}{3} u^2 \right]~, \eea etc. They are orthogonal
under the inner product (no complex conjugation!) \be \langle n |m
\rangle \equiv \int_0^1 \frac{du}{1-u^2} \Psi_n (u) \Psi_m (u) \ee
defined by appropriate analytic continuation of the parameter
$\xi$. To normalize them ($\langle n |n\rangle = 1$), choose \be
A_0^2 = \frac{4\Gamma (-i\xi)}{\sqrt\pi \Gamma (-i\xi -
\frac{3}{2})} \ \ , \ \ \ \ A_1^2 = \frac{6\Gamma
(-i\xi-1)}{\sqrt\pi (-i\xi - \frac{5}{2} ) \Gamma (-i\xi -
\frac{7}{2})}~, \ee etc.

Moving away from the critical point, the frequencies shift by \be
\delta\omega_n = \frac{1}{2\omega_n} \frac{\langle n |\mathcal{H}'
|n\rangle}{\langle n|n\rangle}~, \ee where \be \mathcal{H}' \Psi_n
= -u^2\hat f(u)\left(u^2\hat f(u)\Psi_n'\right)' + \hat V (u)
\Psi_n -\omega^2\Psi_n~, \ee where we applied standard first-order
perturbation theory.

We obtain for the axial modes \bea\label{eqar1} \delta\omega_0 &=&
i \left( 1 - \frac{1}{r_+^2} \right) \left[ \frac{3}{2} +
i\xi+\left( \frac{3(1\pm\Delta)}{4} + i\xi \right) \frac{4\Gamma
(-i\xi)}{\sqrt\pi \Gamma(\frac{3}{2}-i\xi)}
\right]~, \nonumber\\
\delta\omega_1 &=& i\left( 1 - \frac{1}{r_+^2} \right) \left[
\frac{7}{2}+i\xi - \frac{\left\{ \frac{27}{4} (1\pm\Delta) -2 +9
\left( 1 + \frac{1\pm\Delta}{2} \right) i\xi - 14\xi^2 \right\}
\Gamma (-1-i\xi)}{\sqrt\pi \Gamma (\frac{3}{2} - i\xi)}
 \right]\nonumber~.\\ \eea
For small $\xi$, the change in the imaginary part is negligible
whereas the change in the real part is \be \delta\omega_0 \approx
-\left( 1 - \frac{1}{r_+^2} \right)
\frac{3(1\pm\Delta)}{2\pi\xi}~. \ee For $Z_2$ modes, above the
critical point ($r_+>1$), $\delta\omega_0 < 0$ and the real part
decreases. There is a critical value of $\xi$ (determined by
$\xi+\delta\omega_0\approx 0$), \be\label{eqana6a} \xi_0 \approx
\sqrt{12(1+\Delta)(T-T_0)}~, \ee where we used $T-T_0 \approx
\frac{1}{4\pi} \left( 1 - \frac{1}{r_+^2} \right)$, below which
the mode does not propagate (purely dissipative mode). It turns
out that for $\xi < \xi_{0}$
 there is a pair of
purely dissipative modes.

The first harmonic behaves similarly with a higher critical value
of $\xi$, \be\label{eqana6b} \xi_1 \approx
\sqrt{[27(1+\Delta)-8](T-T_0)} \ee below which it turns into a
pair of purely dissipative modes.

Below the critical point ($r_+ <1$), $\delta\omega_0 > 0$ and the
real part of the $Z_2$ modes increases. The modes do not become
purely dissipative for any value of $\xi$.

Also notice that above the critical point, $\delta\omega_n$
increases with $n$, therefore the real part decreases with $n$
(positive slope) whereas below the critical point we obtain a {\em
negative} slope for propagating modes.

$Z_1$ modes behave in the opposite way because $1-\Delta <0$.
Above the critical point, these modes never become purely
dissipative. Below the critical point, we obtain the critical
values \be\label{eqana4} \xi_0 \approx \sqrt{12(\Delta-1)(T-T_0)}
\ \ , \ \ \ \ \xi_1 \approx \sqrt{[ 8 + 27 (\Delta-1) ] (T-T_0)}
\ee similar to $Z_2$ modes above the critical point.

For the polar modes we obtain \bea \delta\omega_0 &=& i \left( 1 -
\frac{1}{r_+^2} \right) \left[ \frac{3}{2} + i\xi+\left(
3(1\pm\Delta)\frac{\Lambda -2}{\Lambda +2} + i\xi \right)
\frac{4\Gamma (-i\xi)}{\sqrt\pi \Gamma(\frac{3}{2}-i\xi)}
\right]~, \nonumber\\
\delta\omega_1 &=& i\left( 1 - \frac{1}{r_+^2} \right) \left[
\frac{7}{2}+i\xi - \frac{\left\{ -\frac{9}{2}
(1\pm\Delta)\frac{\Lambda + \frac{3}{2}}{\Lambda + 2} \left(
\frac{3}{2} + i\xi \right) -2 +9 i\xi - 14\xi^2 \right\} \Gamma
(-1-i\xi)}{\sqrt\pi \Gamma (\frac{3}{2} - i\xi)}
 \right]~.\nonumber\\ \eea
These modes coincide with their axial counterparts for $\Delta =
-1$ (electromagnetic modes).

\subsection{Uncharged Black Holes}

In the case of no charge in the black hole, the wave equations
simplify because $u^2\hat f(u)$ has at most three distinct roots,
$u = 1, \eta, -\frac{\eta}{1+\eta}$, where \be \eta = -
\frac{2}{1+ \sqrt{1-4(1-1/r_+^2)}}~. \ee Then the solution to the
wave equation may be written in terms of a Heun function.

For $Z_1$ perturbations, the potentials for axial and polar modes
coincide reducing to the electromagnetic potential. The
wavefunction may be written as \bea \Psi (u) &=&
(u-\eta)^{-i\hat\omega \frac{\eta}{(1-\eta)(2+\eta)}} \left(
u+\frac{\eta}{1+\eta} \right)^{i\hat\omega \frac{\eta
(1+\eta)}{(2+\eta)(1+2\eta)}}
(1-u)^{i\hat\omega\frac{\eta^2}{(1-\eta)(1+2\eta)}}
\nonumber\\
& & \times\ \mathrm{Heun} (a,q,\alpha,\beta,\gamma,\delta,
\frac{1-u}{1-\eta})~. \eea The Heun function obeys the equation
\be z(z-1)(z-a)F'' -
[(-\alpha-\beta-1)z^2+((\delta+\gamma)a-\delta+\alpha+\beta+1)z-a\gamma]
F' - [-\alpha\beta z+q] F = 0 \ee and the various constants are
\[ a = \frac{1+2\eta}{1-\eta^2} \ \ , \ \ \ \
q = - \Lambda \frac{1+\eta +\eta^2}{1-\eta^2}~, \]
\[ \alpha = 0 \ \ , \ \ \ \ \beta = 2 \ \ , \ \ \ \
\gamma = \frac{1+\eta -2(1-i\hat\omega)\eta^2}{(1-\eta)(1+2\eta)}
\ \ , \ \ \ \ \delta = \frac{2 - (2i\hat\omega +1)\eta -
\eta^2}{(1-\eta) (2+\eta)}~.
\]
It behaves nicely at the horizon ($u\to 1$). Requiring $\Psi (0) =
0$ yields the constraint \be\label{eq54} \mathrm{Heun}
(a,q,\alpha,\beta,\gamma,\delta, \frac{1}{1-\eta}) = 0~, \ee which
may be solved for $\hat\omega$ to obtain the quasi-normal
frequencies of axial $Z_1$ modes.

As $r_+\to\infty$, the two lowest purely dissipative modes
asymptote respectively to \be\label{eqana2} \omega_0 =
-i\frac{3}{2} r_+ \ \ , \ \ \ \omega_1 = - 3i r_+ \ee if $\Lambda$
is kept constant. As $\Lambda$ increases, the two modes approach
each other coalescing at \be\label{eqana3} \Lambda = 0.115 r_+^2 \
\ , \ \ \ \ \omega_0 = \omega_1 = -2.05 i r_+ \ee Beyond this
point, they develop a finite real part.

For axial $Z_2$ perturbations, the solution to the wave equation
may be similarly written as \bea \Psi (u) &=&
(u-\eta)^{-i\hat\omega \frac{\eta}{(1-\eta)(2+\eta)}} \left(
u+\frac{\eta}{1+\eta} \right)^{i\hat\omega \frac{\eta
(1+\eta)}{(2+\eta)(1+2\eta)}} (1-u)^{i\hat\omega
\frac{\eta^2}{(1-\eta)(1+2\eta)}}
\nonumber\\
&& \times\ \mathrm{Heun} (a,q',\alpha',\beta',\gamma,\delta,
\frac{1-u}{1-\eta})~, \eea where the parameters $a,\gamma,\delta$
are same as before and \be q' = - \frac{(1+\eta +\eta^2) \Lambda +
3(1+\eta)}{1-\eta^2} \ \ , \ \ \ \ \alpha' = -1 \ \ , \ \ \ \
\beta' = 3~. \ee It behaves nicely at the horizon ($u\to 1$).
Requiring $\Psi(0) = 0$ yields the constraint \be\label{eq57}
\mathrm{Heun} (a,q',\alpha',\beta',\gamma,\delta,
\frac{1}{1-\eta}) = 0~, \ee which may be solved for $\hat\omega$
to obtain the quasi-normal frequencies of axial $Z_2$ modes.

As $r_+\to\infty$, we find a single purely dissipative axial $Z_2$
mode which asymptotes to \be \omega_0 = -i \frac{\Lambda +2}{3r_+}
\ee confirming the earlier analytical result (\ref{eqZ2ana}) in
the case $q=0$.

The above results do not apply to polar $Z_2$ perturbations due to
the additional singularity in the potential which survives the
limit $q\to 0$.

\section{Numerical Calculations and Results}
\label{sec4}
 We
briefly review the method of Horowitz and Hubeny \cite{HH} as it
is applied to our problem.
After performing the transformation $\Psi(r)=\psi_\omega(r) e^{-i
\omega r_*},$ the wave equation (\ref{sch}) becomes \be f(r)
\fr{d^2\psi_\omega(r)}{d r^2}+ \left(\fr{d f(r)}{d r} -2 i \omega
\right) \fr{d \psi_\omega(r)}{d r} =V(r)
 \psi_\omega(r)~,\ee where the potential $V(r)$ is given by
  (\ref{eq15}) or (\ref{eq15p}). The
change of variables $r=1/x$  yields an equation of the form
$$s(x) \left[(x-x_+)^2 \fr{d^2 \psi_\omega(x)}{d x^2}\right] +t(x)
\left[(x-x_+) \fr{d \psi_\omega(x)}{d x}\right]+ u(x)
\psi_\omega(x)=0~,$$ where $x_+ = 1/r_+$
and $s(x), t(x)$ and $u(x)$ are  given by
\begin{eqnarray} s(x) &=& \sum_{n=0}^k s_n(x-x_+)^k~, \nonumber\\
t(x) &=& \sum_{n=0}^k t_n(x-x_+)^k~,\nonumber\\
u(x) &=& \sum_{n=0}^k u_n(x-x_+)^k~,\nonumber\end{eqnarray} where
$k=3$ for axial and $k=7$ for polar perturbations. Expanding the
wavefunction around the (inverse) horizon $x_+$,
\be\psi_\omega(x)=\sum_0^\infty a_n(\omega) (x-x_+)^n~,\ee we
arrive at a recurrence formula for the coefficients, \be
a_n(\omega) = -\fr{1}{n(n-1)s_0+n t_0+u_0}\sum_{m=n-3}^{n-1} [m
(m-1) s_{n-m} +m t_{n-m} +u_{n-m}] a_m(\omega)~.\ee We note that
the few coefficients $a_m(\omega)$ with negative index $m$ which
will appear for $n < 2$ should be set to zero, while $a_0(\omega)$
is set to one. Since the wave function should vanish at infinity
$(r \rightarrow \infty, x=0),$ we deduce
\be \psi_\omega(0) \equiv \sum_0^\infty a_n(\omega)
(-x_+)^n=0~.\label{HH} \ee
The solutions of this equation are precisely the quasi-normal
frequencies.
\subsection{Overview and Spacing}

\begin{figure}[!t]
\centering
\includegraphics[angle=-90,scale=0.5]{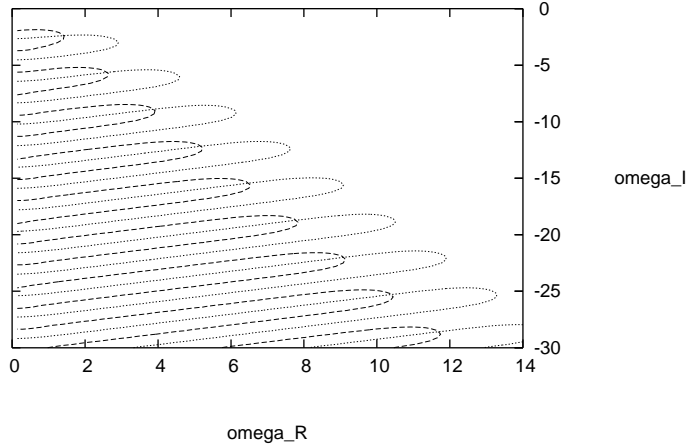}
\caption{QNMs of axial $Z_1$ perturbations with $r_+=1.50,~
\nu=0.10.$} \label{rp=150_010_z1}
\end{figure}

To begin with, let us recall the expression for the temperature
given earlier: $T = \frac{3r^{4}_{+}-r^{2}_{+}-q^{2}}{4 \pi
r^{3}_{+}}.$ The constraint that $T$ should be positive yields the
inequality $q^2 \le q^2_{max} \equiv 3 r_+^4-r_+^2,$ so the charge
may be expressed through the parameter \be \nu \equiv
\frac{Q}{Q_{max}} \ \ , \ \ \ \ Q_{max} =
\frac{\sigma}{4\pi}\sqrt{3 r_+^4-r_+^2}~.\ee We note that this
parameter is related to the charge-to-mass ratio $\lambda$ through
the equation $$\lambda^2=\frac{4 \nu^2 (3 r_+^2-1)}{(3 \nu^2 r_+^2
+r_+^2-\nu^2-1)^2}.$$ One needs to examine separately large
$(r_+>1)$ and small $(r_+<1)$ horizons.  A typical graph for axial
perturbations at $r_+=1.50, \nu=0.10 \ (\lambda=0.37)$ may be seen
in Fig.~\ref{rp=150_010_z1}. The figure depicts the curves
$\Re[\psi(\omega)]=0$ and $\Im[\psi(\omega)]=0$ in the complex
$\omega$ plane. The QNMs are given by the intersections of the
curves. We remark that, if one views $\omega_I$ versus $\omega_R,$
the slope is negative, as can be seen in Fig.~\ref{rp=150_010_z1};
this feature does not change if one varies $\nu.$ This specific
graph depicts axial $Z_1$ QNMs at $r_+=1.50$ and $\nu=0.10,$ but
no qualitative change occurs if one considers $Z_2$ rather than
$Z_1$ perturbations. The behaviour is similar for polar
perturbations: for this value of $r_+$ only some quantitative
changes have been observed between the polar perturbations and
their axial counterparts.

\begin{figure}[!t]
\centering
\includegraphics[angle=-90,scale=0.5]{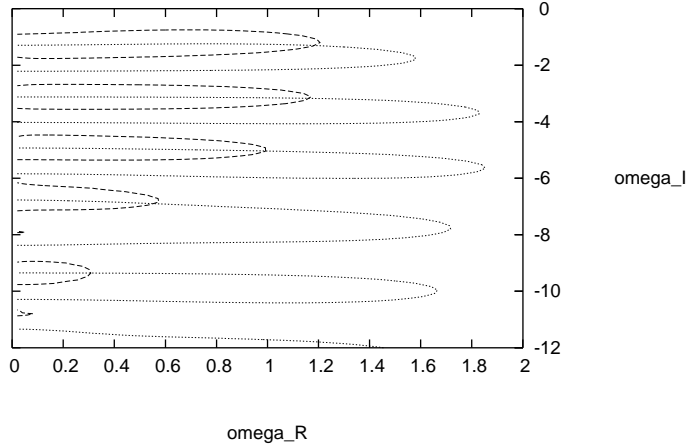}
\caption{QNMs of axial $Z_2$ perturbations with $r_+=0.95,~
\nu=0.10.$} \label{rp=095_010_z1}
\end{figure}

However, for the small horizon $r_+=0.95$ the behaviour of QNMs is
different. For $\nu=0.10 \ (\lambda=3.25)$ and axial $Z_2$
perturbations (Fig.~\ref{rp=095_010_z1}) we find a finite number
of propagating QNMs with a positive slope. (We note that there
might also exist purely dissipative modes with vanishing real
part, which are not clearly visible on such graphs; we will
examine them in the next section).

\begin{figure}[h]
\centering
\includegraphics[angle=-90,scale=0.5]{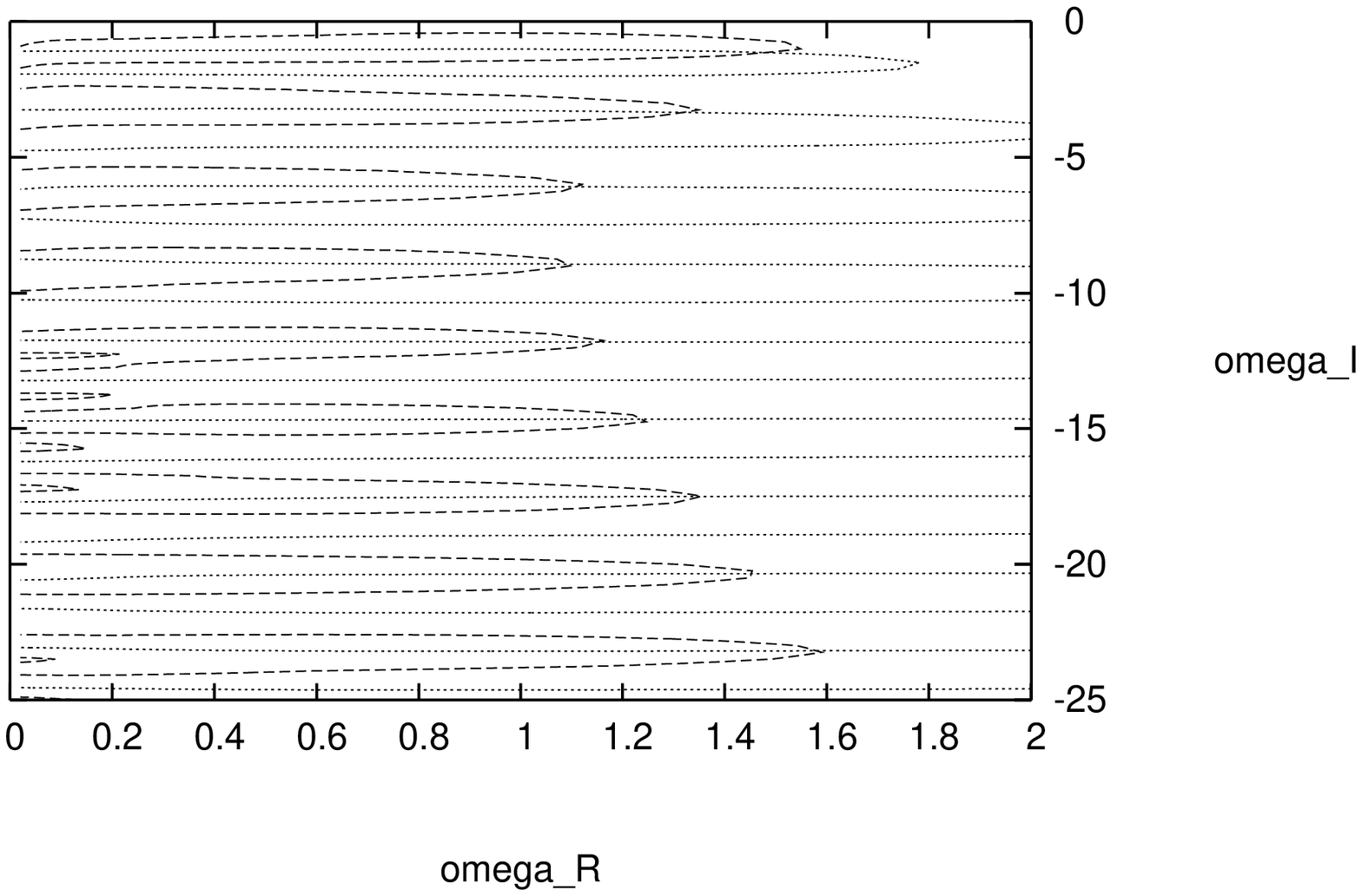}
\caption{QNMs for axial $Z_2$ perturbations with
$r_+=0.95,~\nu=0.40.$} \label{rp=095_040_z2}
\end{figure}

For $\nu=0.40 \ (\lambda=5.95)$ the pattern is qualitatively
different as can be seen in Fig.~\ref{rp=095_040_z2}: a part with
positive slope coexists with the negative slope frequencies and
the number of QNMs is infinite again. We find similar results for
the axial $Z_1$ perturbations, as well as for the polar $Z_1$ and
$Z_2$ perturbations.

We may also make some quantitative statements about these modes.
The asymptotic spacing between the QNMs may be easily read off
from the figures and the results are presented in Table
\ref{Table2}.
For $r_+=0.95$, the spacing is not constant; thus we report the
difference between the lowest and the second lowest QNMs just to
get some feeling for the order of magnitude.
An important observation is that positive and negative slope parts
coexist for $r_+=0.95$ when $\nu$ is large enough
(Fig.~\ref{rp=095_040_z2}), so the results for the imaginary parts
refer rather to the absolute value than to values with a well
defined sign all the way.
\begin{table}
\begin{center}
\begin{tabular}{||c|c|c|c|c||}\hline $r_+$ & $\nu$ & type & $Z_1$ & $Z_2$
\\ \hline $1.50$ & $0.10$ & axial & $1.25 - 3.25 i$ &
$1.25 - 3.30 i$ \\ \hline $1.50$ & $0.30$ &
axial & $1.05 - 3.85 i$ & $1.05 - 3.85 i$ \\
\hline $1.50$ & $0.00$ & polar & $1.30 - 3.40 i$ & $1.36 - 3.30 i$
\\ \hline $1.50$ & $0.40$ & polar & $0.94 - 4.20
i$ & $1.03 - 4.17 i$ \\ \hline $0.95$ & $0.10$ &
axial & $0.26 + 1.67 i$ & $0.09 + 2.01 i$ \\
\hline $0.95$ & $0.40$ & axial & $0.00 + 1.80 i$ & $0.17 + 2.20 i$
\\ \hline $0.95$ & $0.00$ & polar & $0.32 + 1.80
i$ & $0.38 + 1.70 i$ \\ \hline $0.95$ & $0.40$ &
polar & $1.90 + 2.30 i$ & $1.42 + 2.43 i$ \\
\hline
\end{tabular}
\end{center}
\caption{\label{Table2} Spacing of QNMs for various values of the
parameters $r_+$
 and $\nu$ with $\xi = 1$.}
\end{table}

\subsection{Lowest Modes}

For $r_+>1$ the QNMs of the form presented in the previous section
may be represented as a multiple of the spacing plus their lowest
possible value, referred to as the offset in the literature. It is
interesting to examine the behaviour of these lowest modes, in
addition to the spacing. The charge is given as a fraction $\nu$
of its maximal value and notice that we cannot increase the charge
parameter $\nu$ beyond a value about $0.40,$ due to convergence
problems.

\subsubsection{Axial Perturbations}

\begin{figure}[h]
\centering
\includegraphics[angle=-90,scale=0.3]{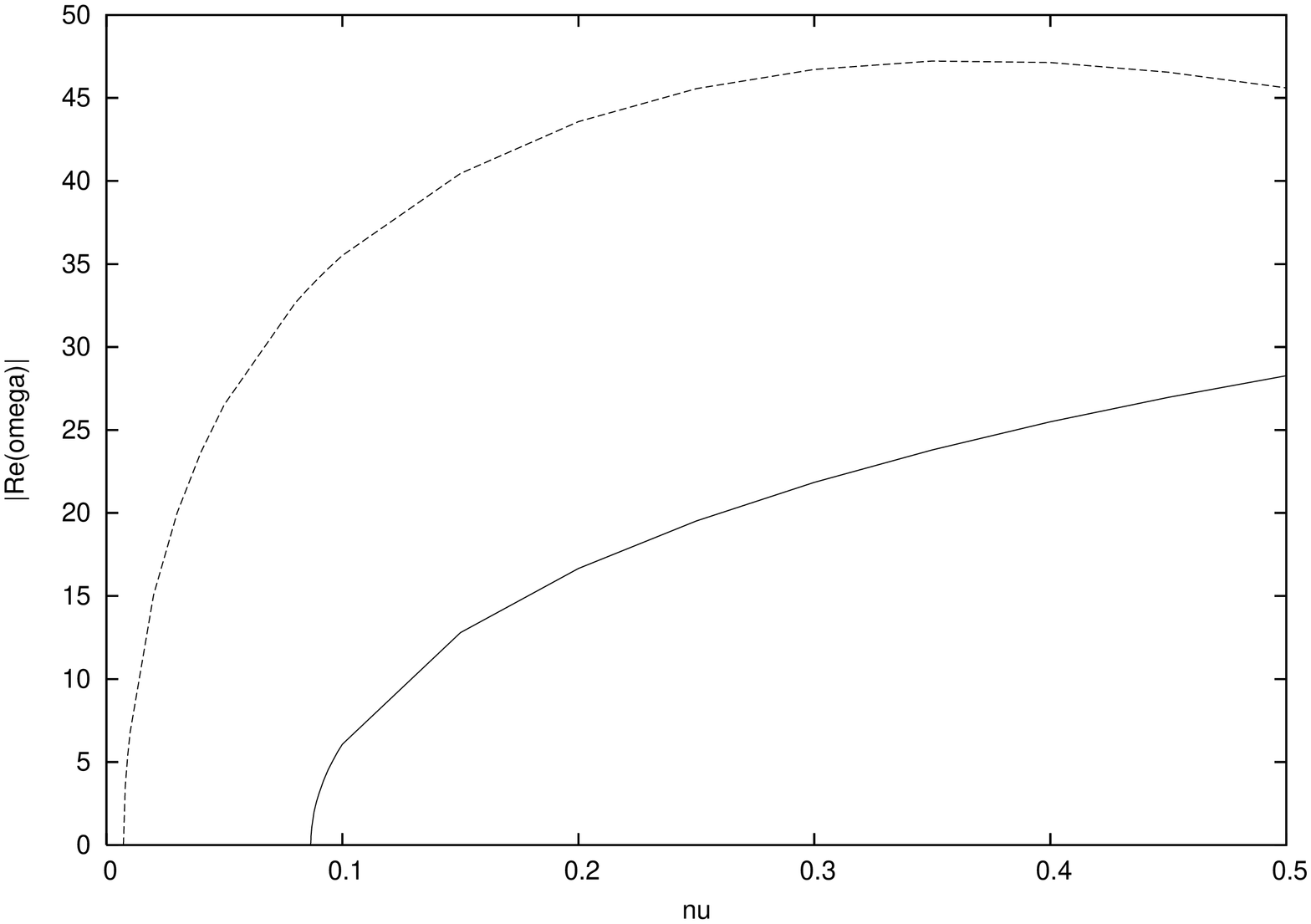}
\caption{The (absolute) real part of the axial $Z_1$ QNMs with
$r_+=20.00$ versus fractional charge $\nu$. The corresponding
$\lambda$ ranges from $0$ to $0.05.$ The solid line corresponds to
the real part of the lowest $Z_1$ mode; the dashed line is the
real part of the second lowest $Z_1$ mode.} \label{ev20.0R}
\end{figure}

We examine the intermediate horizon $r_+=1.50$ and a typical large
horizon, namely $r_+=20.00.$ It turns out that there are
qualitative differences between the two. In Fig.~\ref{ev20.0R} one
may see the real parts of the axial $Z_1$ QNMs for $r_+=20.00$ for
large horizons and small values of the charge purely dissipative
modes are present, which is not the case for intermediate
horizons, such as $r_+=1.50.$ We note that no different behaviour
of this kind shows up for $Z_2$ modes.

\begin{figure}[h]
\centering
\includegraphics[angle=-90,scale=0.3]{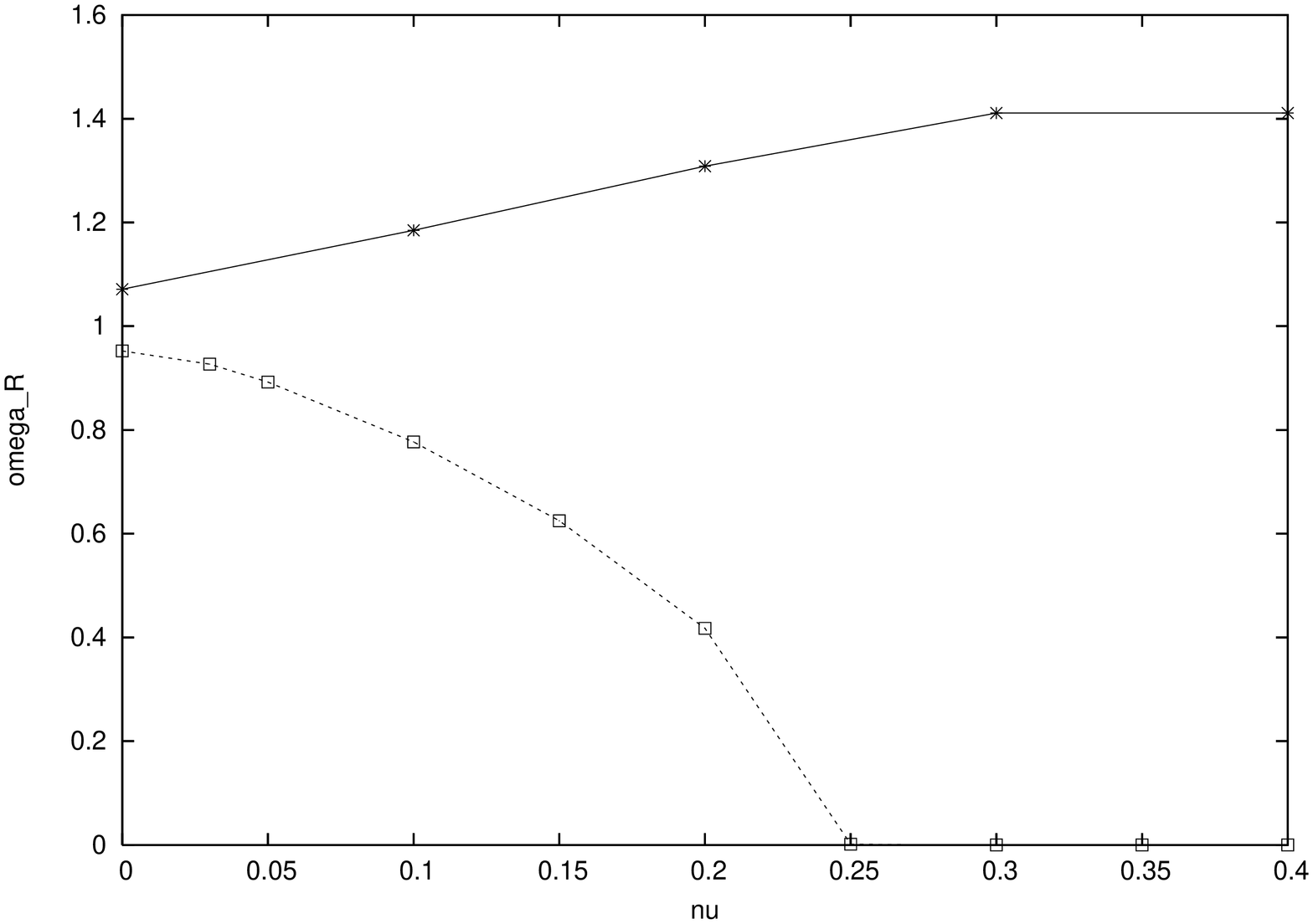}
\caption{The (absolute) real parts of the axial $Z_1$ (lower
curve) and $Z_2$ lowest modes at $r_+=0.95$ versus fractional
charge $\nu$.} \label{095_1_2}
\end{figure}

In Fig.~\ref{095_1_2} we depict the absolute real part of the
lowest axial modes for $r_+=0.95.$ We observe that, as one
increases the charge, $|\Re\omega|$ for $Z_2$ modes approaches a
constant value, which is presumably also the value for the
extremal black holes ($\nu=1$); on the other hand, $|\Re\omega|$
for $Z_1$ modes approaches zero.

\begin{figure}[h]
\centering
\includegraphics[scale=0.3]{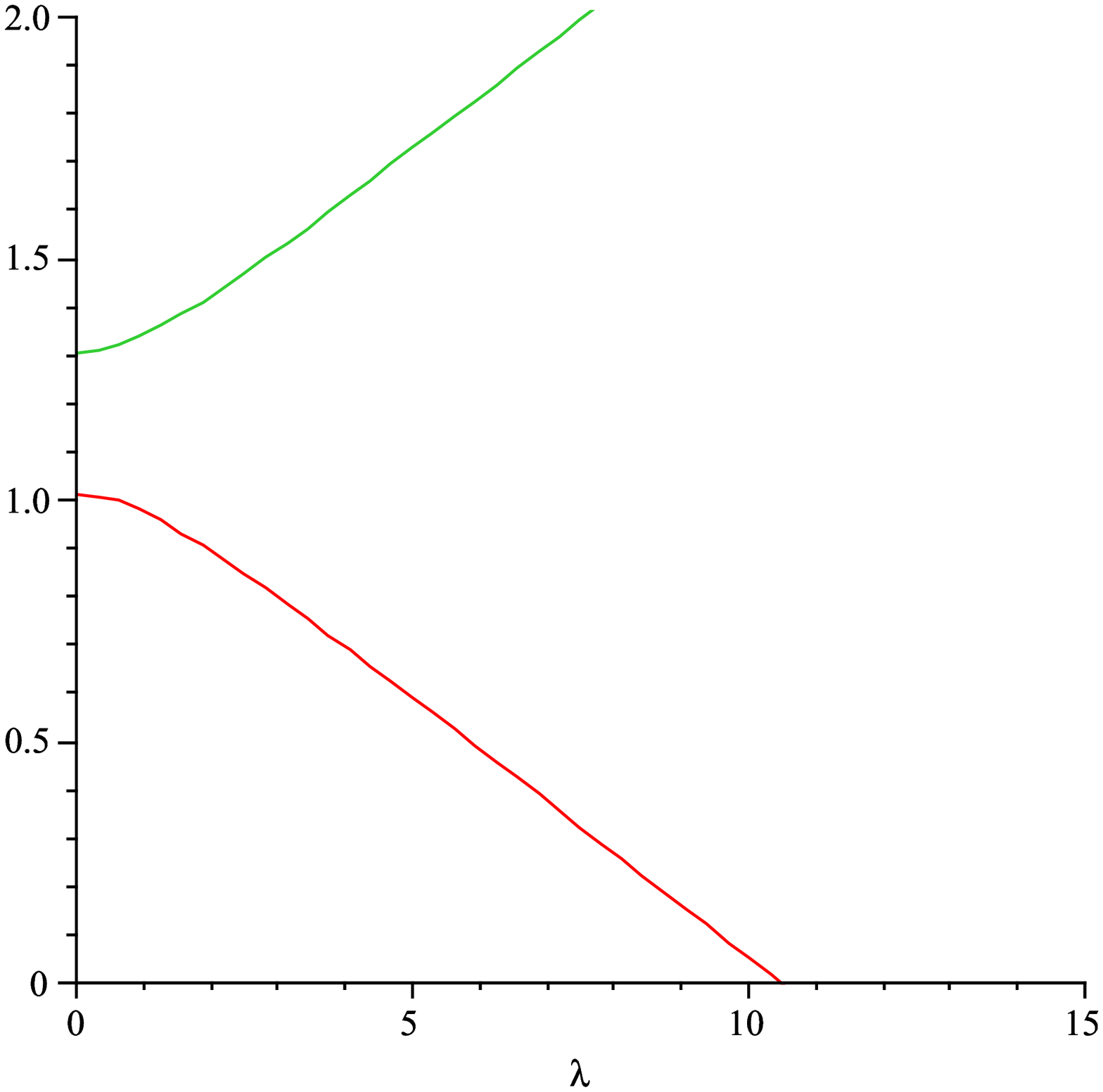}
\includegraphics[scale=0.3]{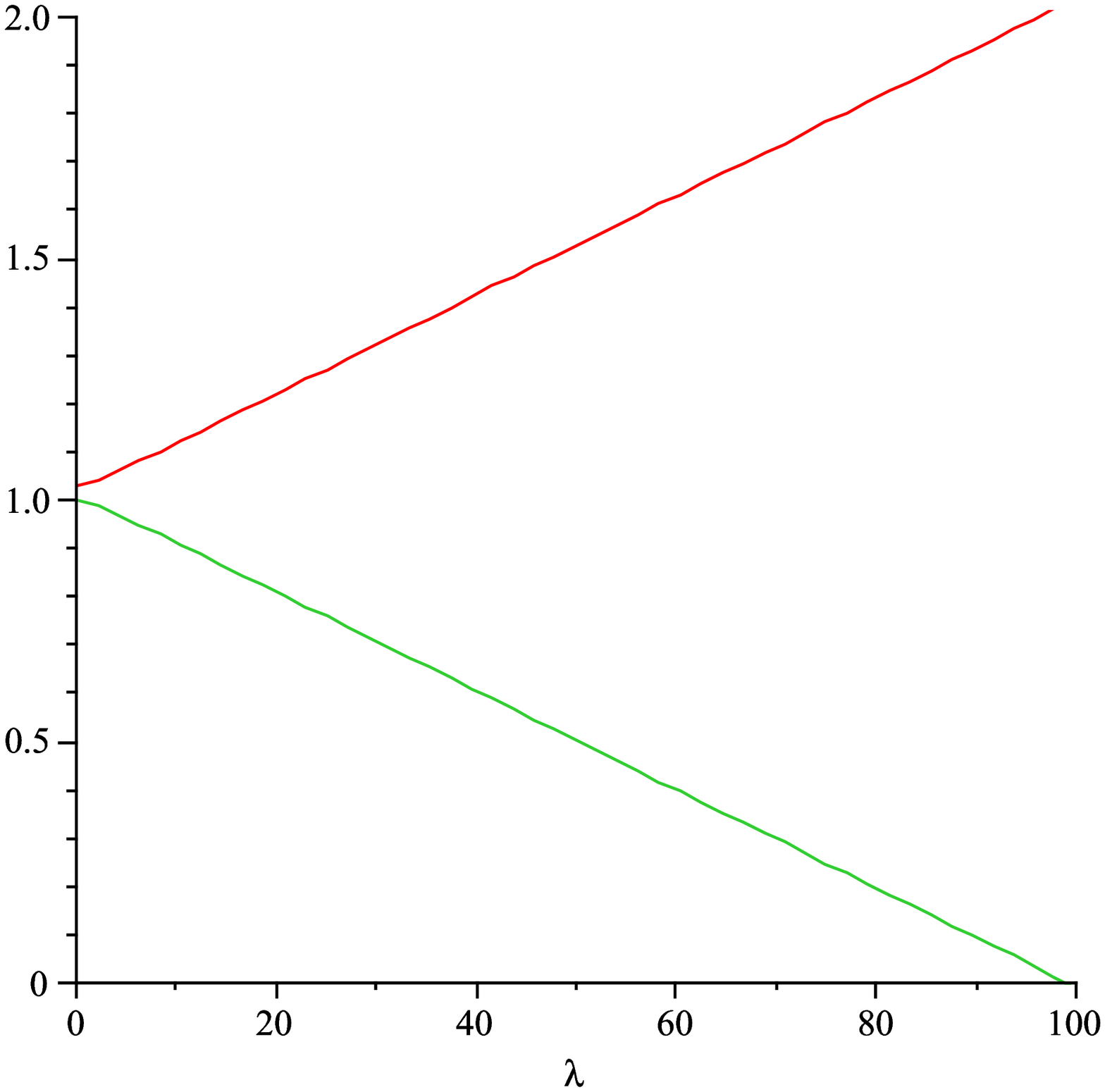}
\caption{The (absolute) real parts of the axial $Z_1$ (lower
curve) and $Z_2$ lowest modes at $r_+=0.95$ and $r_+=0.995$ versus
$\lambda$ using the analytical expressions (\ref{eqana1}) and
(\ref{eqar1}).} \label{095_ana}
\end{figure}

\begin{figure}[!t]
\centering
\includegraphics[scale=0.3]{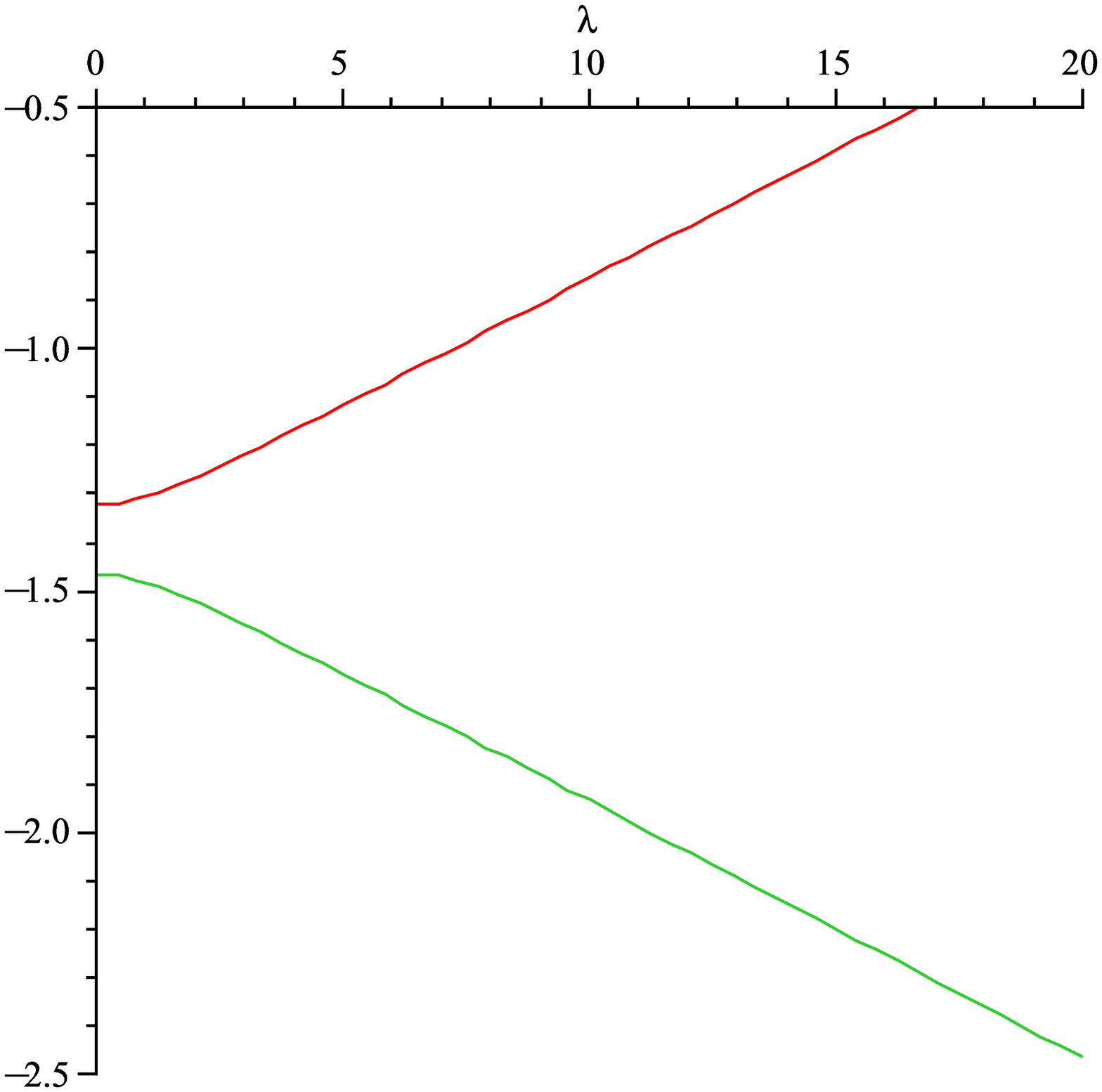}
\includegraphics[scale=0.3]{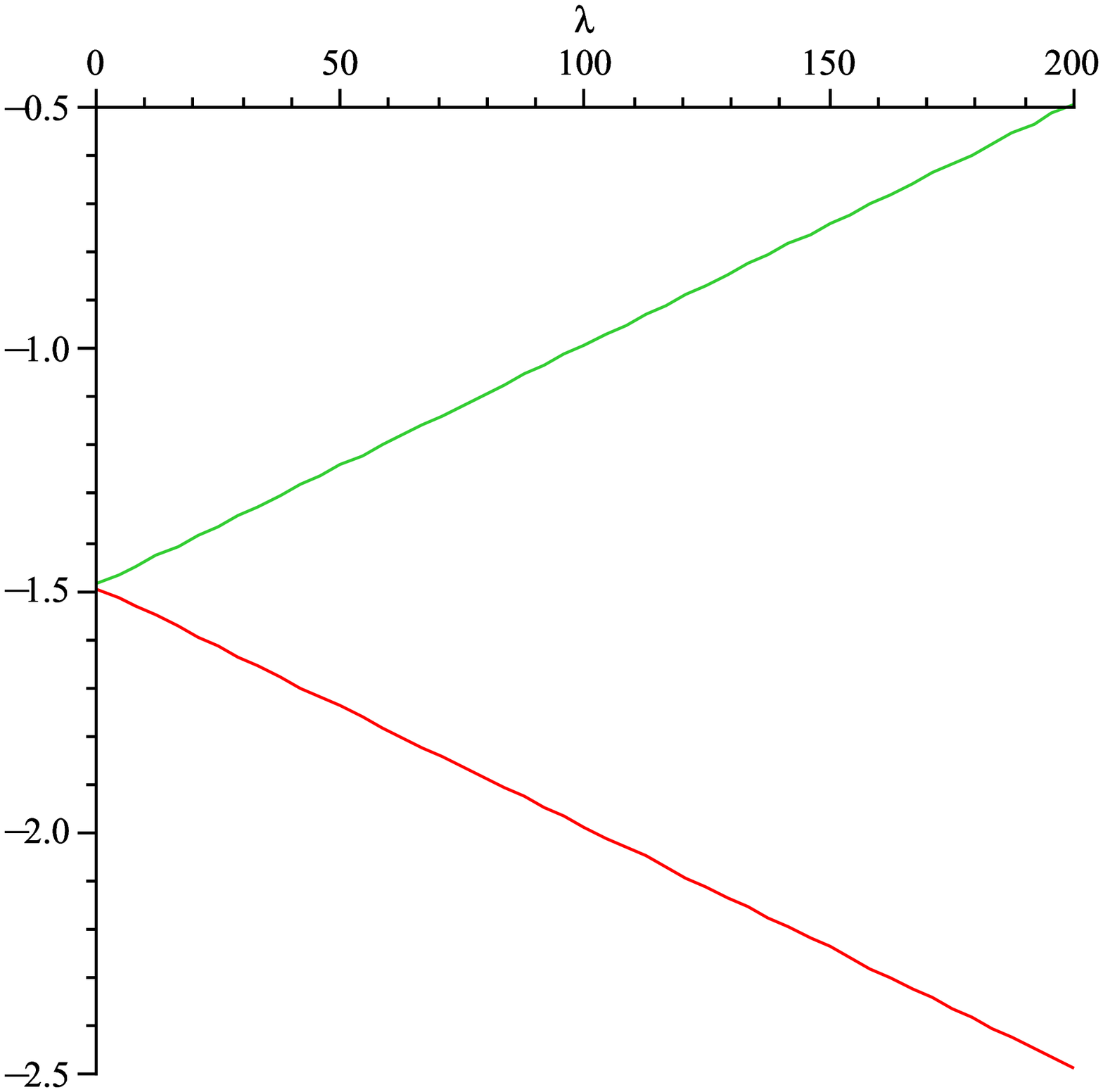}
\caption{The imaginary parts of the axial $Z_1$ (upper curve) and
$Z_2$ lowest modes at $r_+=0.95$ and $r_+=0.995$ versus $\lambda$
using the analytical expressions (\ref{eqana1}) and
(\ref{eqar1}).} \label{095_ana_Im}
\end{figure}

The agreement with the analytical result (\ref{eqar1}) is good
(improving as we approach the critical point $r_+\to 1$). In
Figs.~\ref{095_ana} and \ref{095_ana_Im}, we show plots of the
analytical expressions for $|\Re\omega|$ and $\Im\omega$,
respectively, for $Z_1$ and $Z_2$ axial perturbations for two
different values of the horizon just below the critical point
($r_+ = 0.95$ and $r_+=0.995$). The accuracy of the analytical
approximation decreases as the charge increases and fails when the
first-order correction becomes comparable to the zeroth order
approximation (\ref{eqana1}). Naturally, the point of failure
increases as we approach the critical point.


\subsubsection{Polar Perturbations}

\begin{figure}[!t]
\centering
\begin{tabular}{cc}
\includegraphics[angle=-90,scale=0.3]{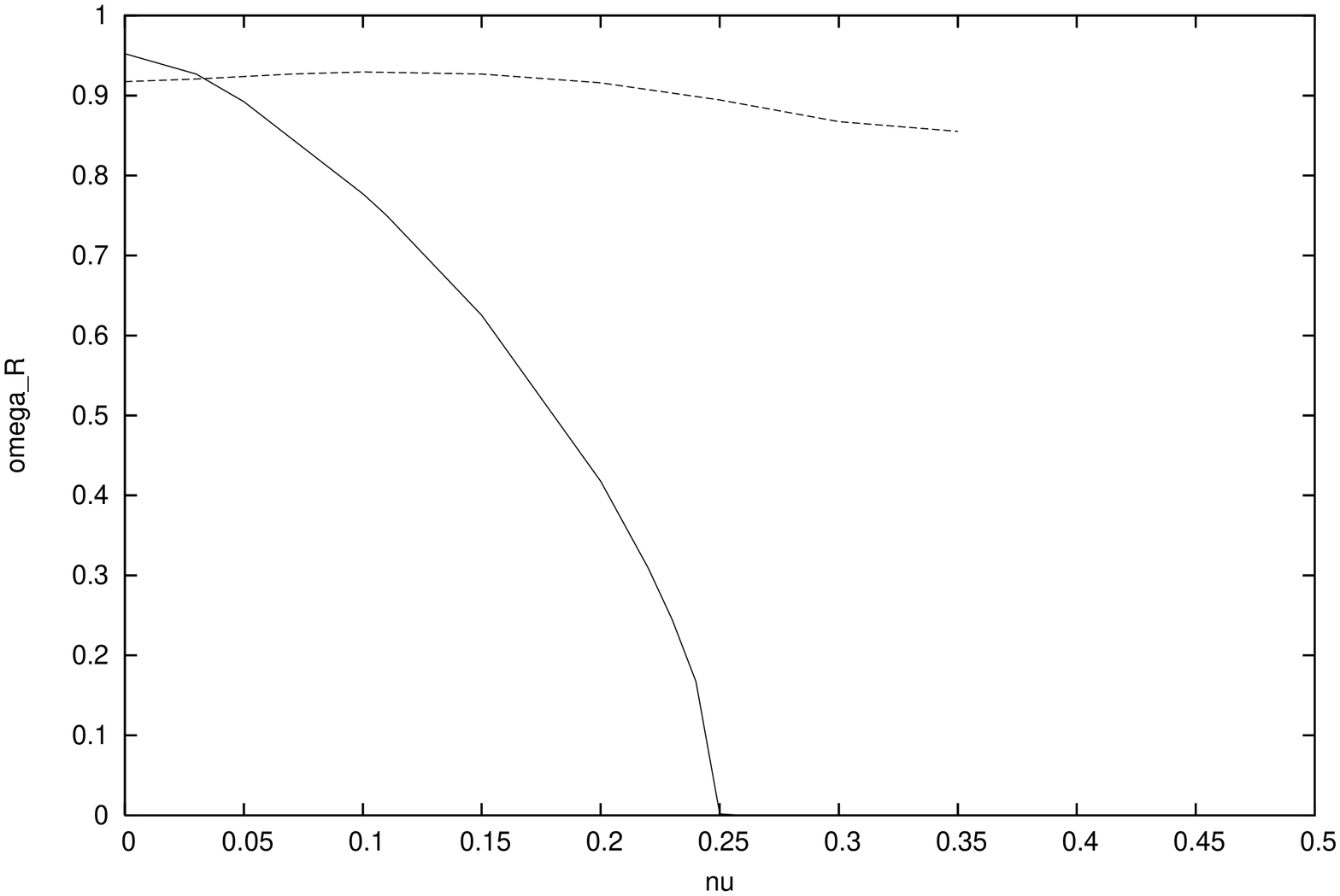}&
\includegraphics[angle=-90,scale=0.3]{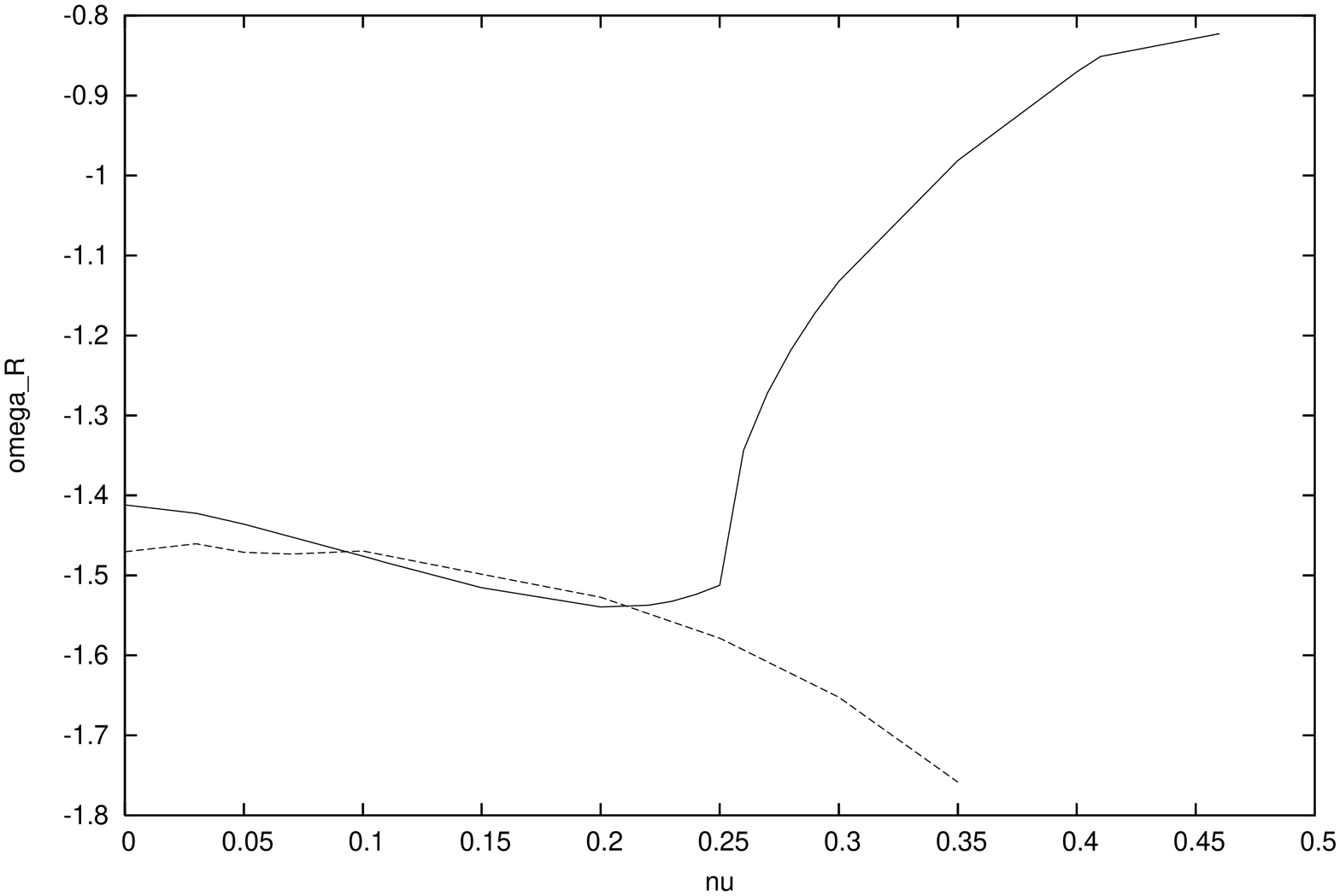}\\ a&b\\
\includegraphics[angle=-90,scale=0.3]{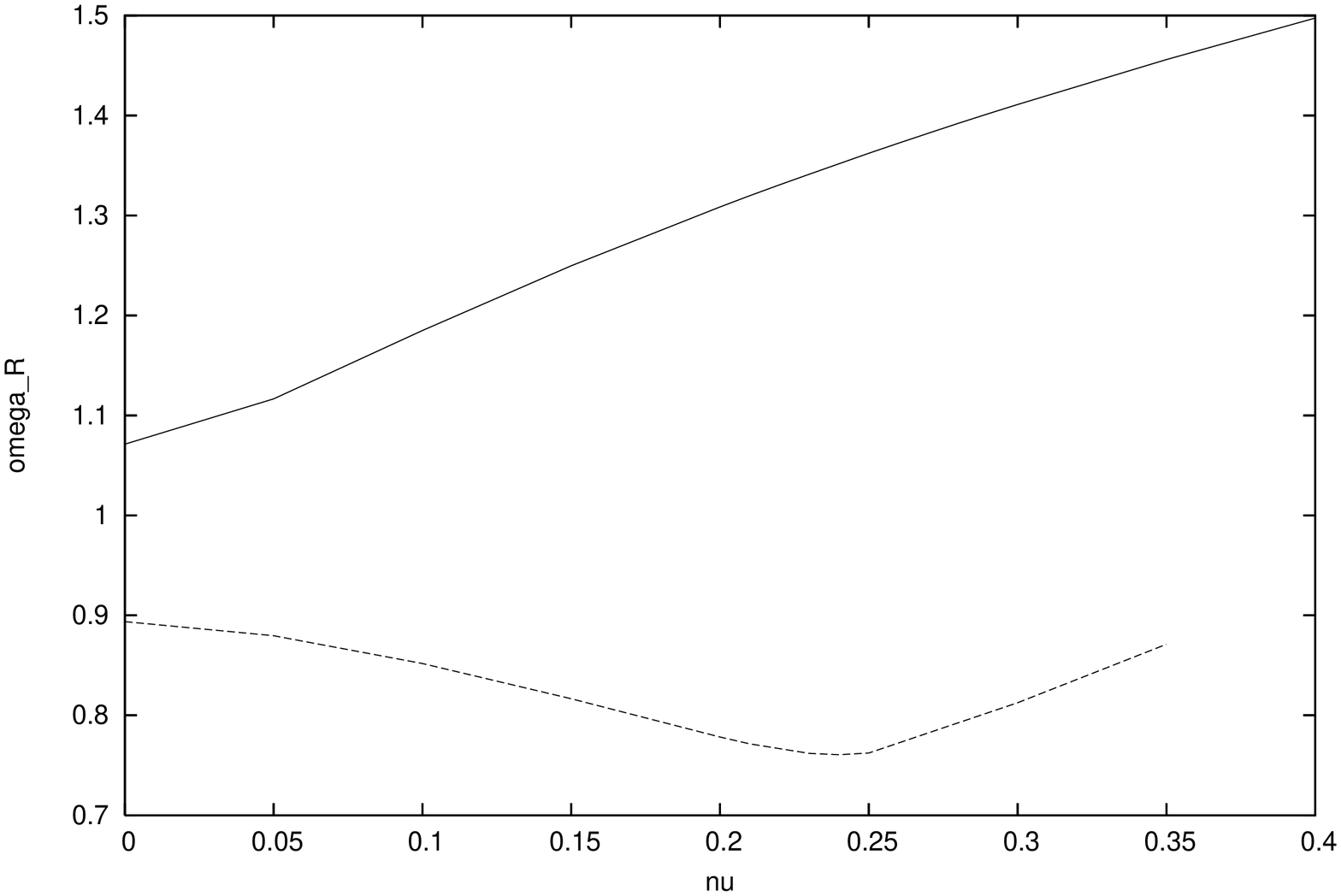}&
\includegraphics[angle=-90,scale=0.3]{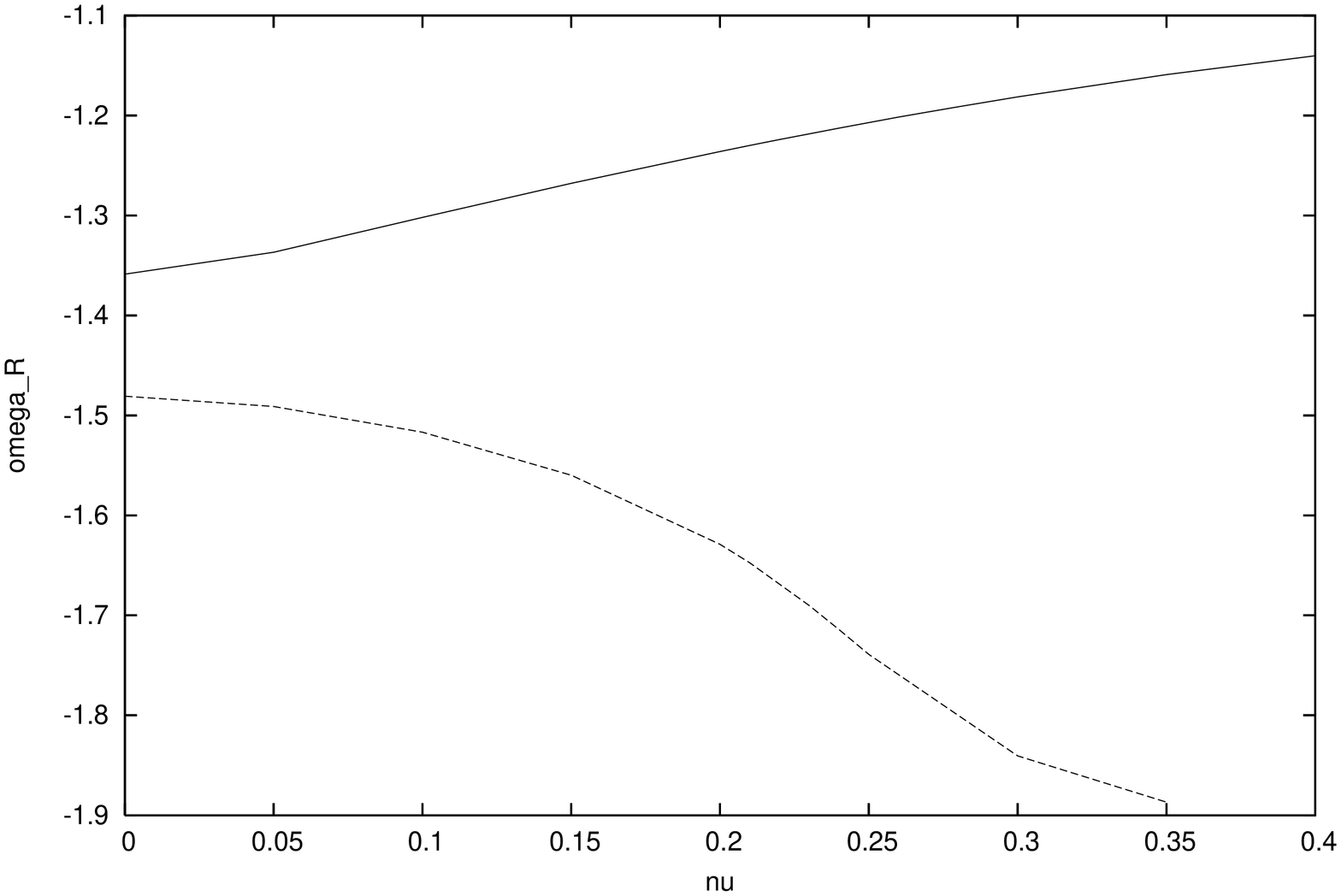}\\ c&d
\end{tabular}
\caption{Panel (a): The real parts of the lowest QNMs of the axial
(lower curve) and polar (upper curve) perturbations for $Z_1$
versus $\nu$ at $r_+=0.95.$ Panel (b): The imaginary parts of the
lowest QNMs of the axial (upper curve) and polar (lower curve)
perturbations for $Z_1$ versus $\nu$ at $r_+=0.95.$ The real,
panel (c) and imaginary, panel (d), parts of the lowest QNMs of
the axial (upper curve) and polar (lower curve) perturbations for
$Z_2$ versus $\nu$ at $r_+=0.95.$} \label{ap1_2}
\end{figure}

Qualitative differences show up for the small value $r_+=0.95,$ so
we concentrate on this case. In Fig.~\ref{ap1_2} we present the
real and imaginary parts of the axial and polar perturbations of
$Z_1$ and $Z_2.$

For $Z_2$ polar perturbations the real parts (Fig.~\ref{ap1_2},
left panel) do not have the striking behaviour of the axial
perturbations, which are zero above $\nu=0.25.$ Similarly the
imaginary parts (right panel) decrease with $\nu$ with no
spectacular change at $\nu=0.25,$ in contrast with their axial
counterparts. Here we also observe the very mild behaviour of the
polar QNMs with the charge as compared with the axial
perturbations.

\section{Purely Dissipative Modes }
\label{sec5}

After the overview of the propagating QNMs presented in the
previous sections, we will make a systematic study of the purely
dissipative modes of electromagnetic and gravitational
perturbations.

\subsection{ Axial $Z_1$ Modes}

\begin{figure}[!t]
\centering
\includegraphics[angle=-90,scale=0.6]{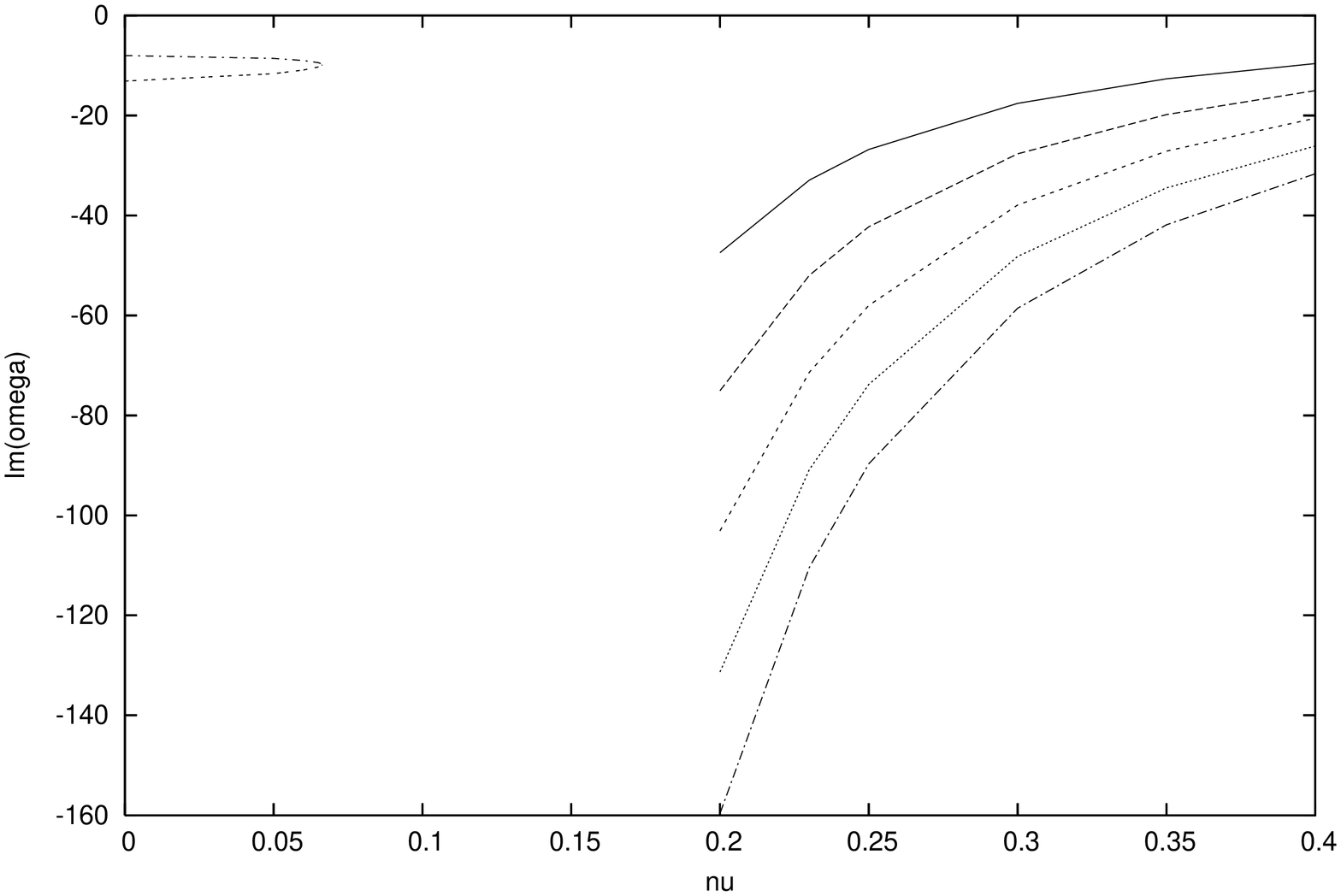}
\caption{The imaginary part of the axial $Z_1$ purely dissipative
QNMs with $r_+=5.00$ (upper left corner), the corresponding
$\lambda$ being smaller than $0.042$ and $r_+=1.50$ (right part of
the figure) versus fractional charge $\nu,$ the corresponding
$\lambda$ ranging between $0.65$ and $0.90.$} \label{150_500}
\end{figure}

Fig.~\ref{150_500} contains the results for purely dissipative
QNMs at $r_+=5.00$ and $r_+=1.50$ versus $\nu.$ For $r_+=5.00$ and
$\nu=0$ there are two purely dissipative modes with the smallest
imaginary parts. For non-zero $\nu$ these modes approach each
other until finally, at $\nu=0.0663$, they take on the same value
and they acquire non-vanishing real parts, transforming into
propagating QNMs. We obtain agreement with the asymptotic analytic
expression (\ref{eqana2}) for $\nu = 0$ ($\Im\omega = -7.50, -15$
for $r_+ = 5.00$).

For intermediate horizons, such as $r_+=1.50,$ the picture is
different: the purely dissipative modes have relatively small
absolute values for $\nu=0.40$ and their absolute values increase
as one approaches $\nu=0.$ We note that for $\nu=0$ there are no
purely dissipative modes at all for this value of $r_+,$ so it
appears that the imaginary parts of the QNMs tend to $-\infty$ as
$\nu \to 0.$ This is in agreement with the modes obtained by
solving the exact analytic equation (\ref{eq54}) which is valid
for $\nu =0$.

The QNMs for $Z_1$ perturbations fall into two disjoint classes;
they belong to either large horizons $(r_+>3.15)$ or small ones
$(r_+<3.15).$ They have different properties for $\nu=0:$ large
horizons have purely imaginary modes in this limit, while small
ones do not. The behaviours of the two classes are qualitatively
different.

\begin{figure}[!b]
\centering
\includegraphics[angle=-90,scale=0.3]{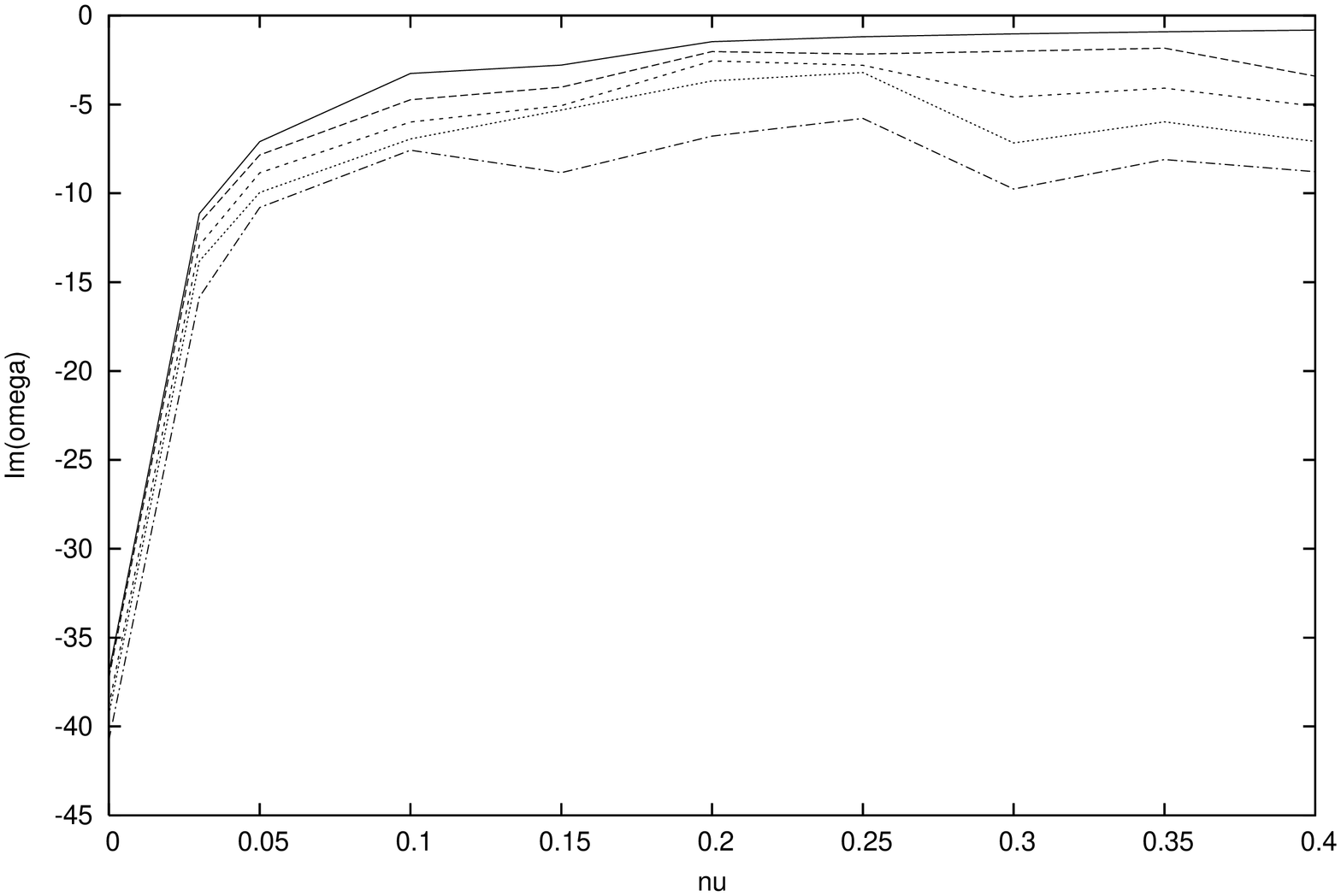}
\caption{$Im(\omega)$ versus $\nu$ for the five lowest $Z_1$
purely dissipative modes at $r_+=0.995.$} \label{z2_995}
\end{figure}

If we go now to a horizon close to and below the critical point,
$r_+=0.995$ (Fig.~\ref{z2_995}), we have in some sense the reverse
behaviour: a finite number of propagating modes corresponds to the
smallest absolute values of the imaginary part, a positive slope
is observed and finally purely dissipative modes appear with large
imaginary parts. For small $\nu$ we observe the phenomenon just
described, the propagating modes appearing last; for somewhat
larger $\nu$ purely dissipative  and propagating QNMs are mixed.
An important point is that the $\nu \to 0$ limit of the imaginary
parts is finite for $r_+=0.995,$ in contrast to what happens for
the intermediate value $r_+=1.50$ (Fig.~\ref{150_500}).

\subsubsection{Temperature Dependence}

\begin{figure}[!t]
\centering
\includegraphics[width=4in]{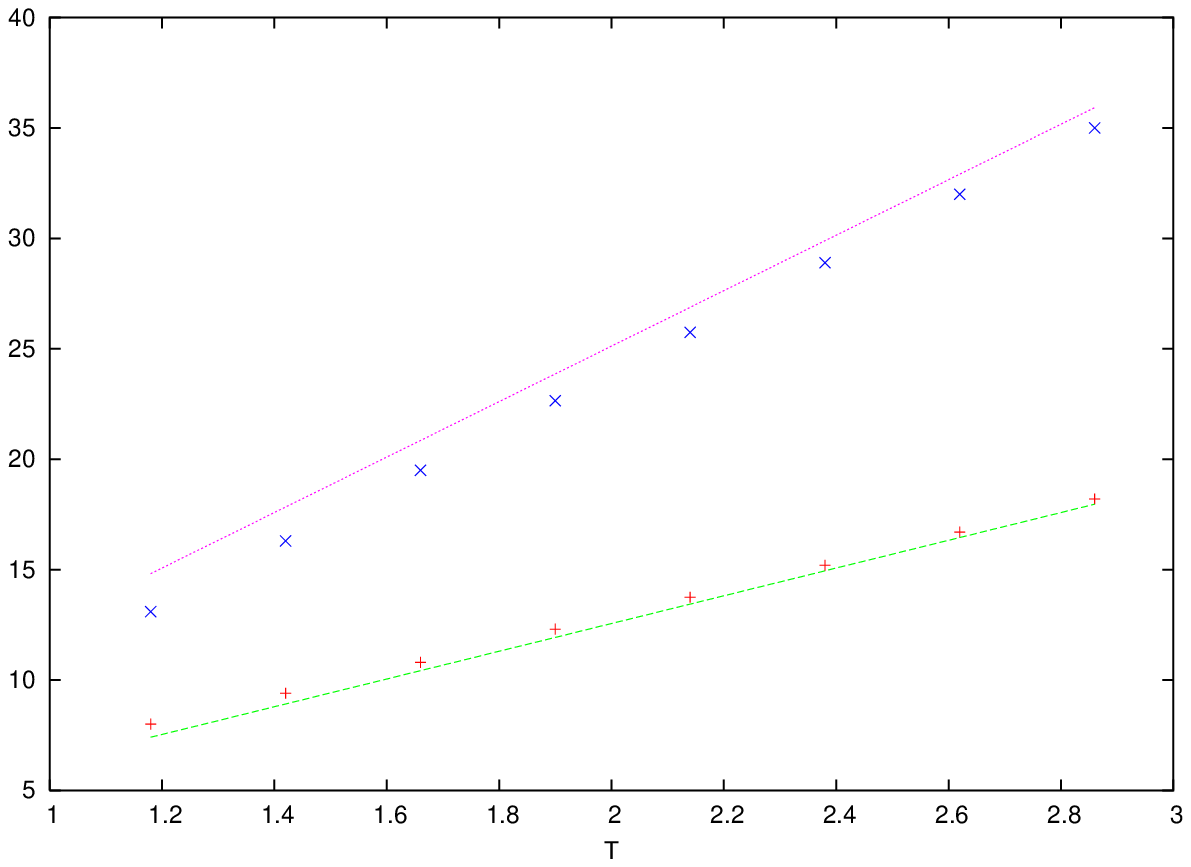}
\caption{The (absolute value of the) imaginary part of the two
$Z_1$ purely dissipative QNMs with $\nu=0$, $\xi =1$ versus
temperature. Horizons vary from $r_+=5.00$ to $r_+=12.00.$ Also
shown are the asymptotic expressions (\ref{eqana2}).}
\label{largeha}
\end{figure}

An issue that should be examined is the $r_+-$dependence of the
imaginary parts. It turns out that it is advantageous to use the
temperature rather than $r_+$ as a variable: in Fig.~\ref{largeha}
we depict the (absolute value of the) imaginary part of the purely
dissipative modes  corresponding to zero charge for large horizons
versus the temperature. We observe that the imaginary parts scale
linearly with the temperature to high accuracy. The data points
have been calculated numerically as well as by solving the exact
analytic equation (\ref{eq54}), the two results being in excellent
agreement with each other. The linear fits agree with the analytic
asymptotic expression (\ref{eqana2}).

\begin{figure}[h]
\centering
\includegraphics[angle=-90,scale=0.4]{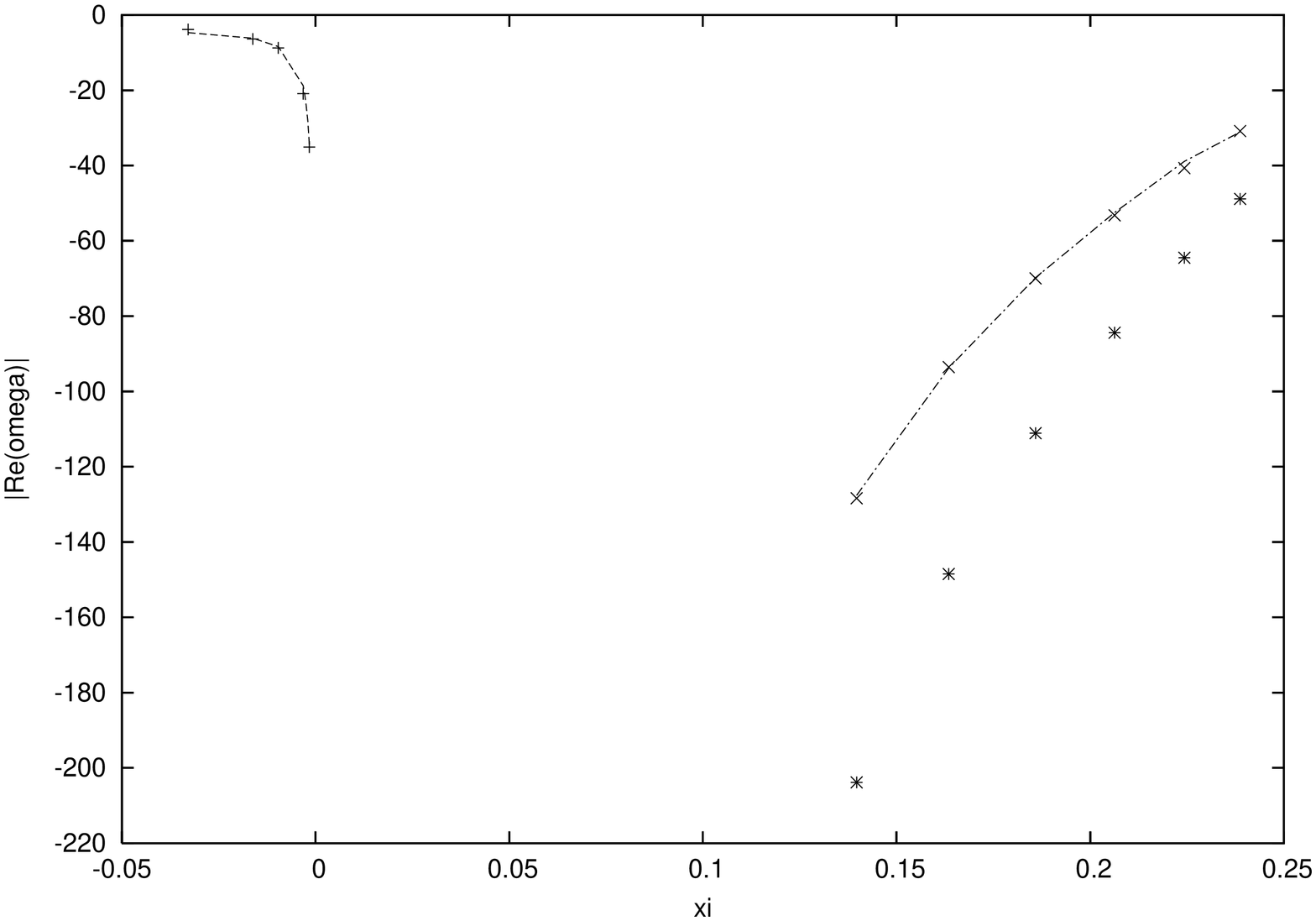}
\caption{The imaginary parts of the two lowest axial $Z_1$ purely
dissipative QNMs with $\lambda=0.50$ versus temperature. For
$T<T_0$ the lowest  modes are represented by points, while the
second lowest ones with lines. For $T>T_0$ the lowest  modes are
represented by points and lines, while the second lowest ones with
points. Horizons vary from $r_+=0.90$ to $r_+=2.00.$}
\label{axial_z1}
\end{figure}

We also examined the intermediate horizons, in particular
$r_+=0.90$ up to $r_+=2.00$ with $\lambda=0.50.$ The relevant
temperature $T-T_0$ ranges from $-0.04$ to $0.24.$ The two lowest
modes are depicted in Fig.~\ref{axial_z1}. Good quality fits
($\chi^2_{d.o.f.} <1)$ of the form $a+\fr{b}{T-T_0}$ are possible
for both branches: $T>T_0$ and $T<T_0.$ The curves for the lowest
modes are also included in the figure.  We observe that there is
an (infinite!) change in the slope of this graph at $T=T_0,$ which
presumably signals a phase transition.

\begin{figure}[!t]
\begin{center}
\includegraphics[angle=0]{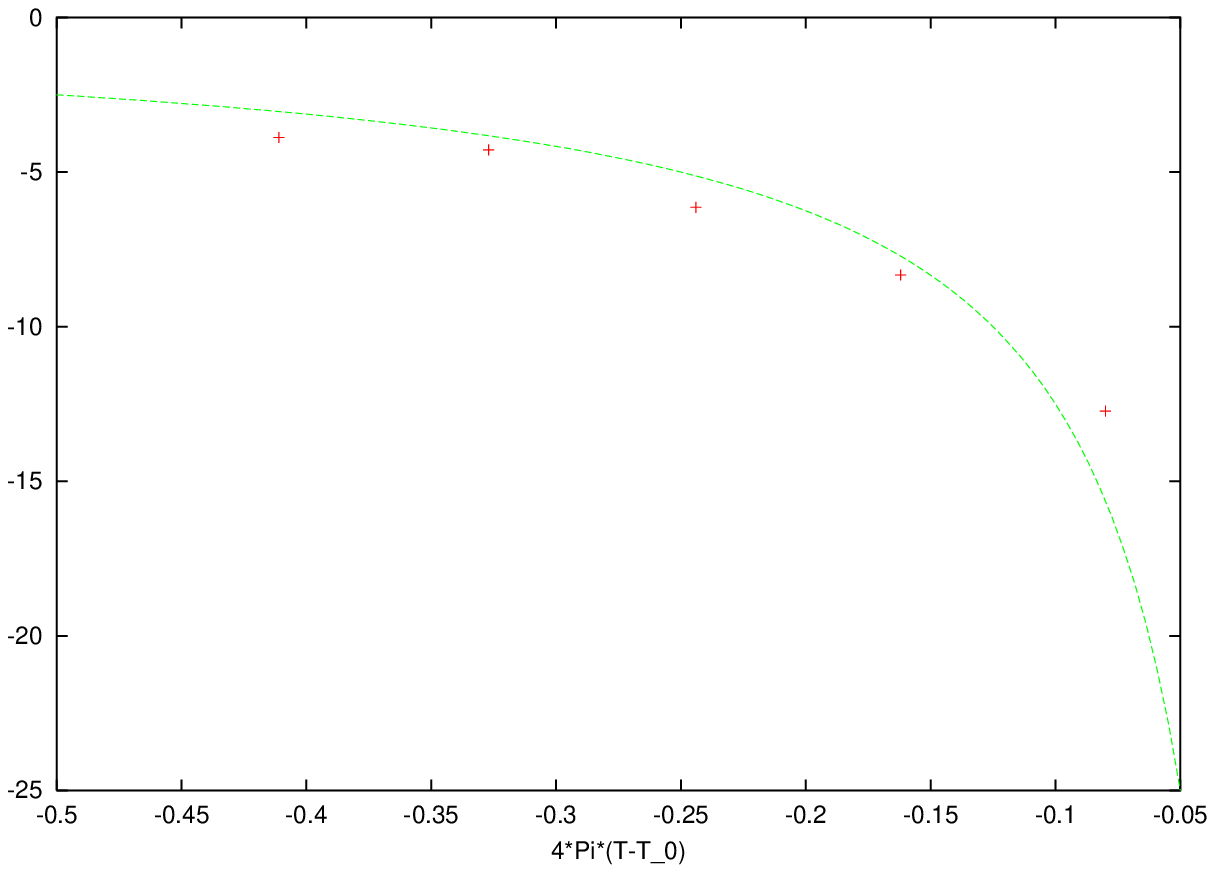}
\end{center}
\caption{The imaginary part of the lowest frequency $\omega_0$
 of the axial $Z_1$ purely dissipative mode versus temperature for $\lambda = 0$, $\xi = 1$ from the exact analytic eq.~(\ref{eqana2}).
  Also shown is a fit $\Im\omega_0 = \frac{1.25}{4\pi (T-T_0)}$. \label{anafig3}}
\end{figure}

In Fig.~\ref{anafig3}, we show a similar plot in the case of no
charge but this time using the exact analytic eq.~(\ref{eqana2}).
We obtain a similar singular behaviour ($\Im\omega \sim
(T-T_0)^{-1}$) reinforcing the conclusion that a phase transition
occurs at $T=T_0$.


\subsubsection{$\xi-$Dependence}

\begin{figure}[!t]
\centering
\includegraphics[angle=-90,scale=0.3]{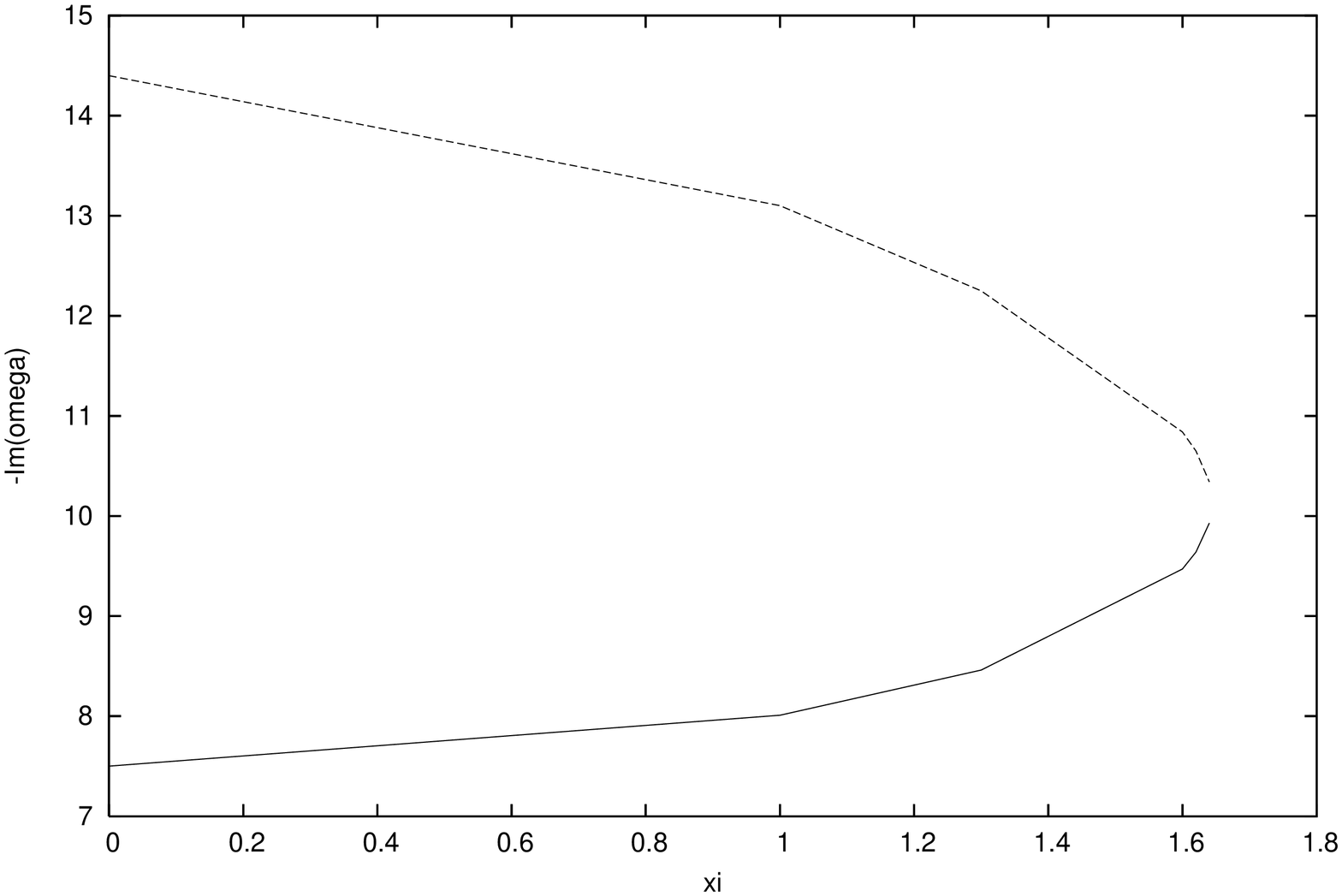}
\caption{The imaginary part of the two $Z_1$ purely dissipative
QNMs for $r_+=5.00$, $\nu =0$ versus $\xi.$ }
\label{largeh_xi_q=0}
\end{figure}

\begin{figure}[!b]
\centering
\includegraphics[angle=-90,scale=0.4]{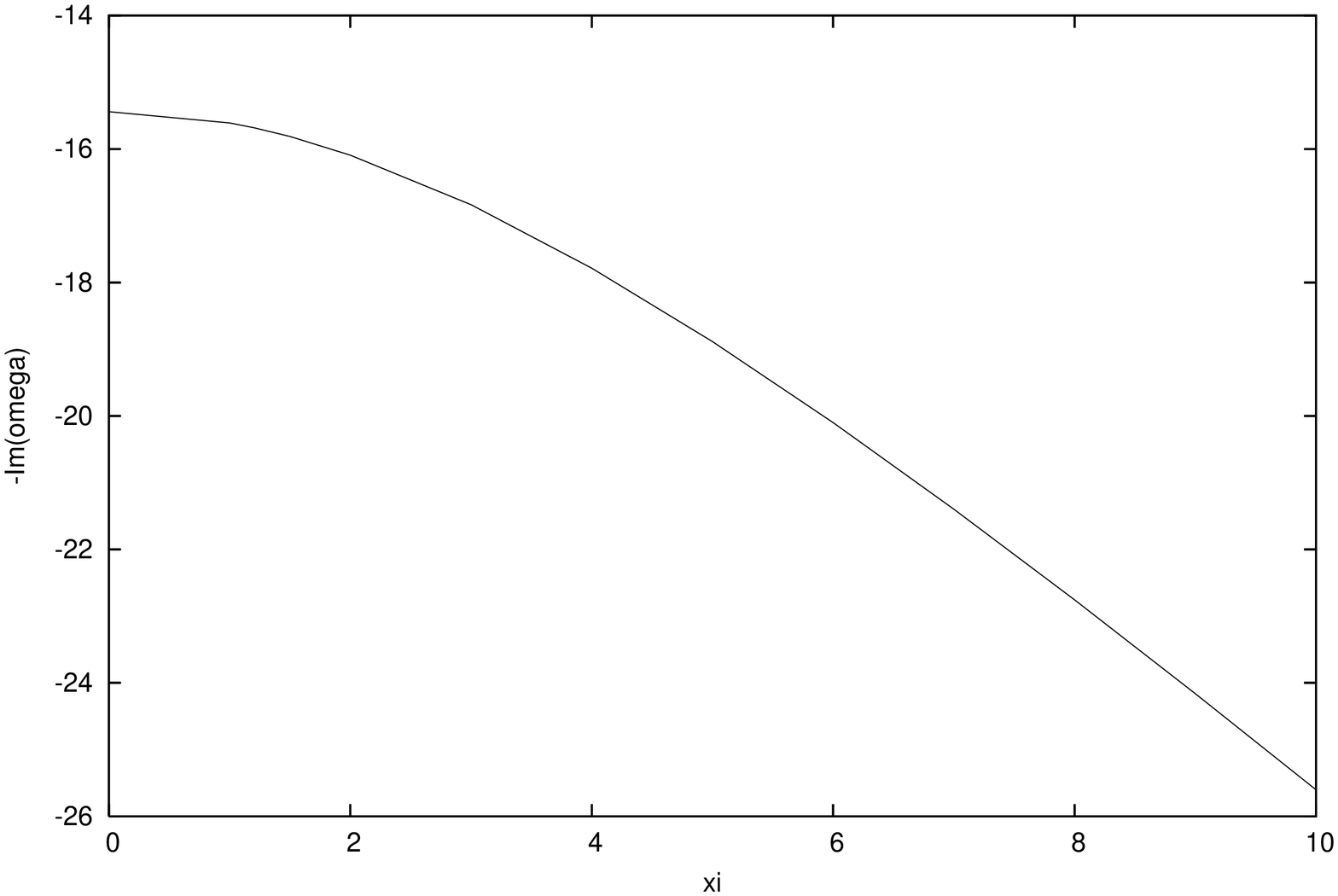}
\caption{The imaginary part of the axial lowest $Z_1$ purely
dissipative QNMs with $r_+=2.00$ and $\nu=0.40$ versus $\xi.$}
\label{intermh_xi_q=040}
\end{figure}

For large horizons (typical value $r_+=5.00$) we consider the
$\nu=0$ case. The $\xi-$dependence is shown in
Fig.~\ref{largeh_xi_q=0}. As $\xi$ grows, the purely dissipative
modes  converge towards each other and finally disappear (that is,
they turn into propagating modes with finite real part), a
behaviour strongly reminiscent of the $\nu-$dependence, depicted
in Fig.~\ref{150_500}. We have chosen to depict in
Fig.~\ref{intermh_xi_q=040} the $\xi-$dependence of the lowest QNM
for a typical intermediate horizon. Numerical results are in
excellent agreement with the results obtained by solving the exact
analytic equation (\ref{eq54}). In particular, for $\xi = 0$, we
obtain the two values $\Im\omega = -15, -7.50$ from
eq.~(\ref{eqana2}) whereas the two modes coalesce at the value
$\Im\omega = -10.25$ at $\Lambda = 2.875$ corresponding to $\xi =
1.62$ (eq.~(\ref{eqana3})), all in agreement with numerical
results.

\begin{figure}[!t]
\begin{center}
\includegraphics[angle=-90,scale=0.3]{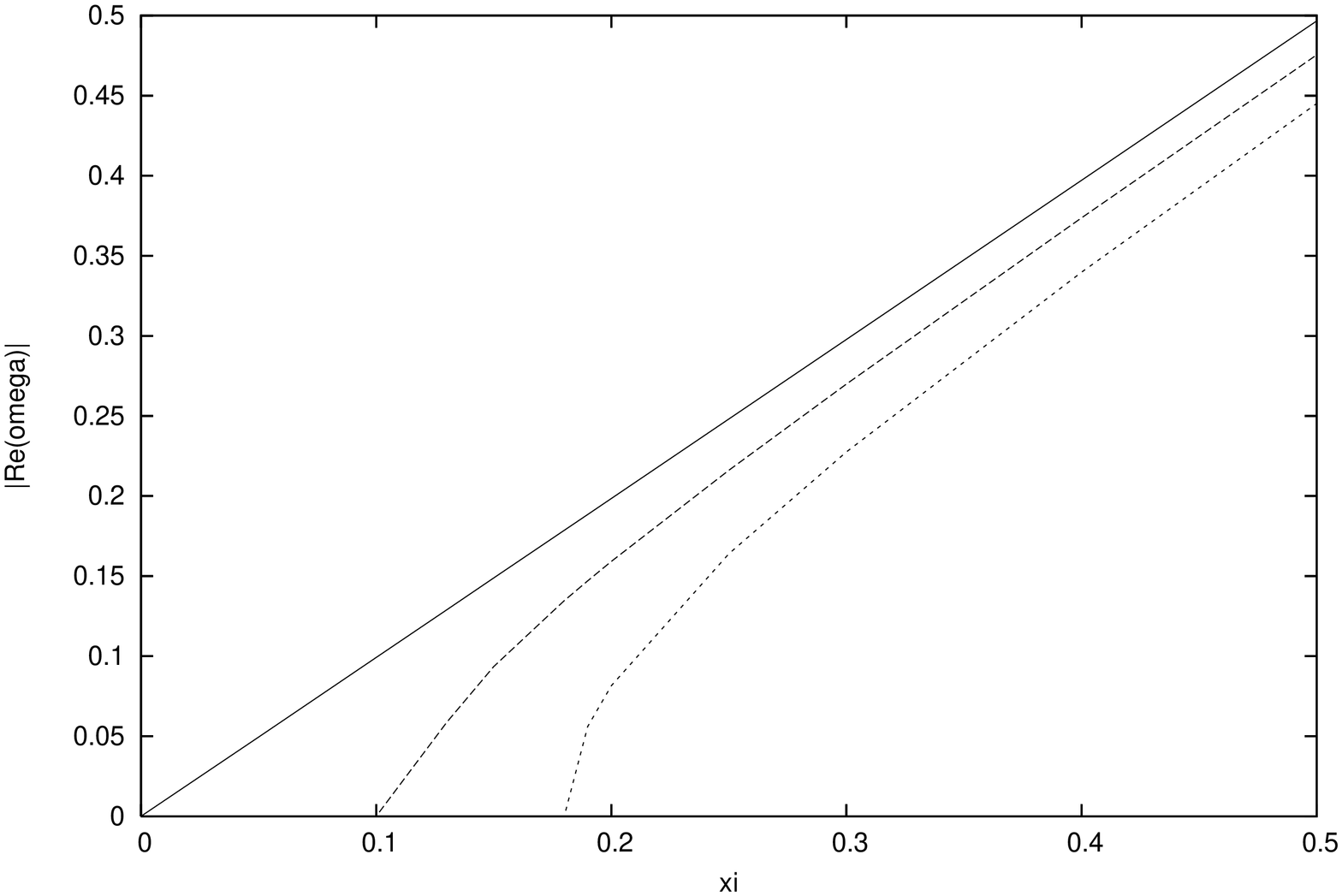}

\includegraphics[height=2in,width=3in]{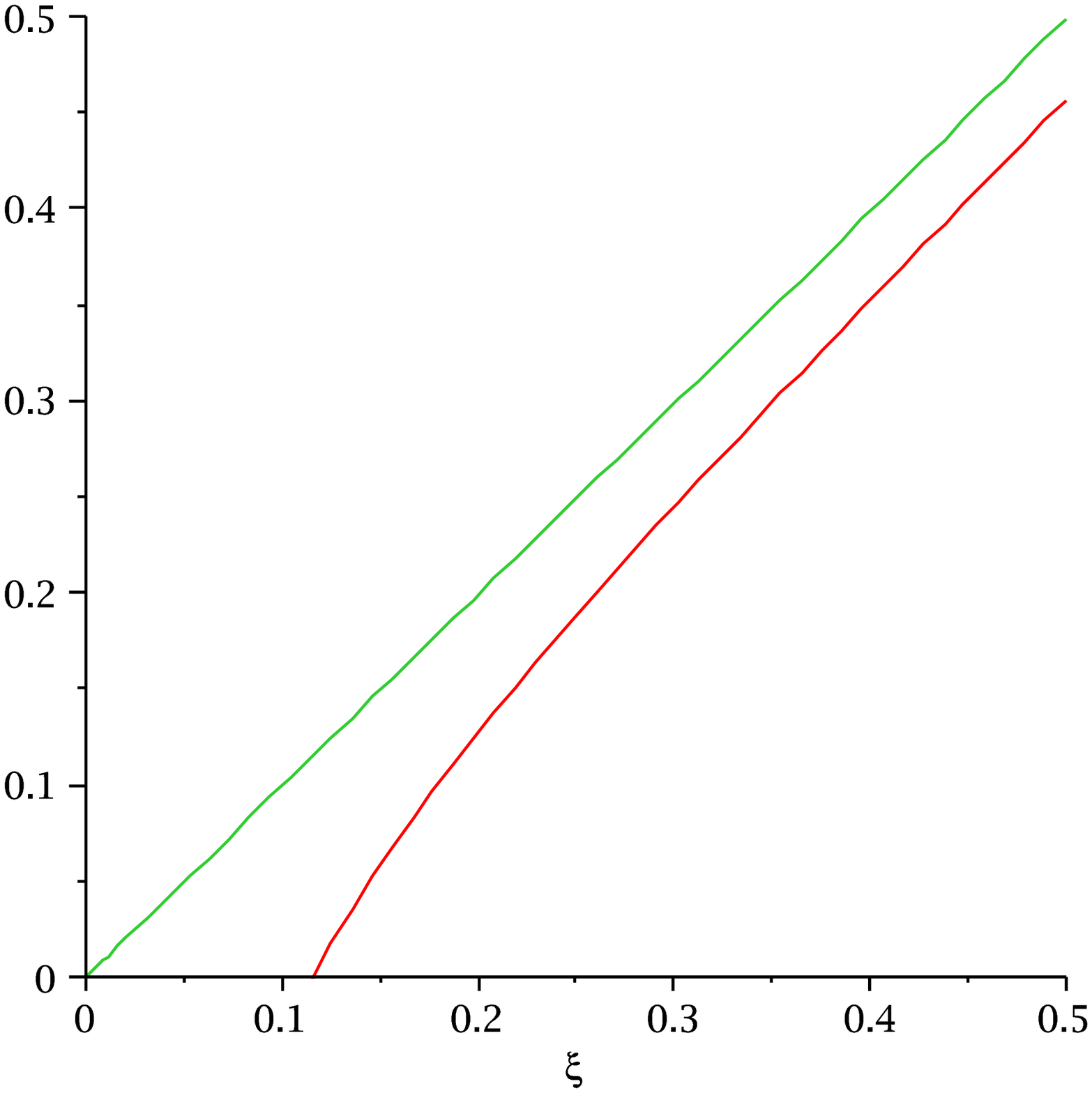}
\end{center}
\caption{The real part of the four lowest axial $Z_1$ QNMs with
$r_+=0.995$ versus the $\xi$ parameter for $\nu=0.00.$ The
imaginary part decreases from left to right.
The second panel shows the analytical estimate for the lowest two
modes (eq.~(\ref{eqar1})) for comparison.} \label{z2xiplt}
\end{figure}

For $r_+=0.995, \ \nu=0$ we know from the analytical calculations
that the imaginary part of the QNMs changes little, so we focus on
the $\xi-$dependence of their real parts. We confirm numerically
that the change in the imaginary parts is small and calculate
numerically the real parts. The results for the lowest QNMs are
shown in Fig.~\ref{z2xiplt}. Both numerical and analytical results
(eq.~(\ref{eqar1})) are shown for comparison.

\begin{figure}[!t]
\begin{center}
\includegraphics[angle=0]{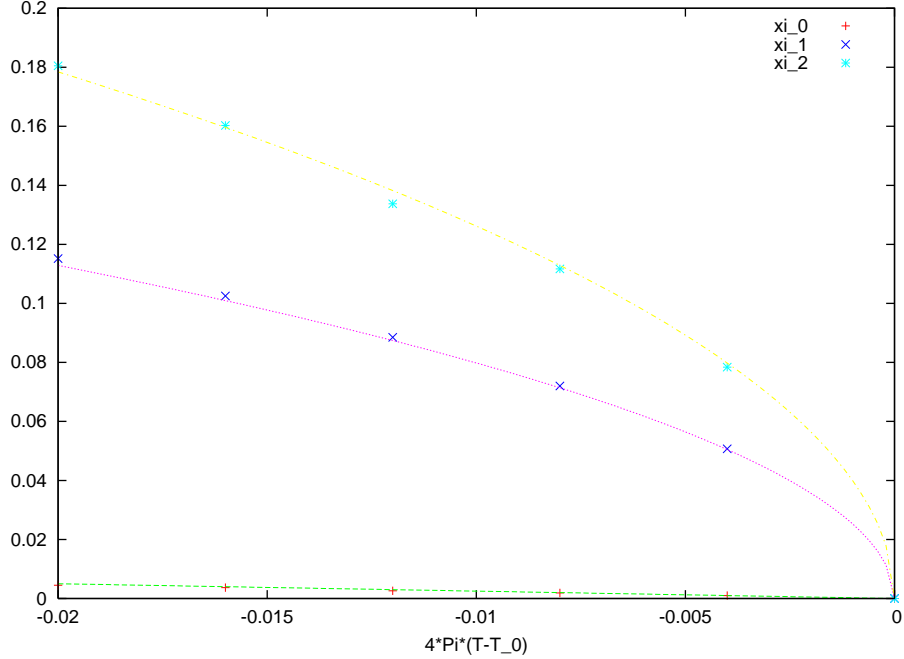}
\end{center}
\caption{The first three critical values of $\xi$ for axial $Z_1$
modes versus temperature for $\nu = 0$. Also shown are fits,
$\xi_0 = \pi (T_0-T)$, $\xi_1 = \sqrt{8 (T_0-T)}$, $\xi_2 =
\sqrt{20 (T_0-T)}$. \label{anafig1}}
\end{figure}
\begin{figure}[!b]
\centering
\includegraphics[angle=-90,scale=0.3]{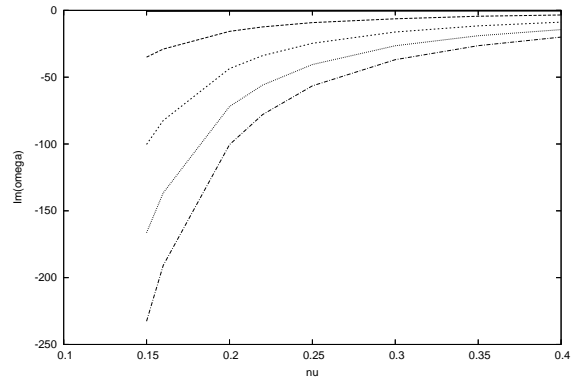}
\caption{$\Im\omega$ versus fractional charge $\nu$ for the five
lowest $Z_2$ purely dissipative modes at $r_+=1.50.$}
\label{z2_150}
\end{figure}
\begin{figure}[!t]
\centering
\includegraphics[angle=-90,scale=0.3]{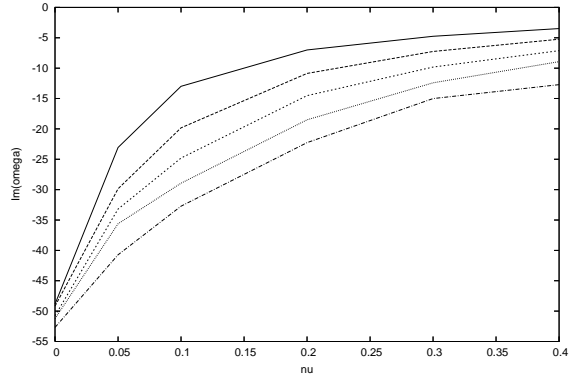}
\caption{The imaginary part of the axial $Z_2$ purely dissipative
QNMs at $r_+=0.995$ versus fractional charge $\nu$.} \label{0995}
\end{figure}

\begin{figure}[!b]
\centering
\includegraphics[angle=-90,scale=0.3]{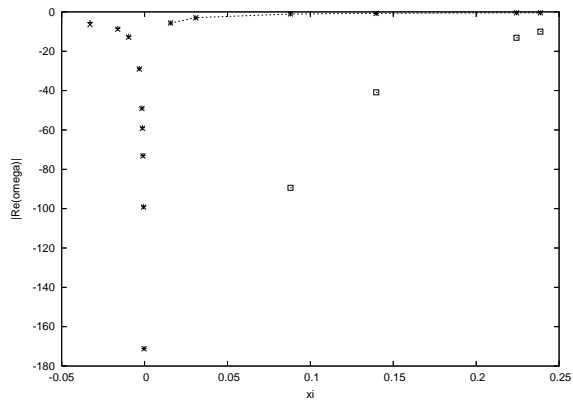}
\caption{$\Im\omega$ versus $T-T_0$ for the lowest and second
lowest axial $Z_2$ modes at $\lambda=0.50.$ For $T<T_0$ the lowest
modes are represented by a line just connecting the points, while
the second lowest ones with points. For $T>T_0$ the lowest modes
are represented by a fit and the points, while the second lowest
ones with points. Horizon values vary from $r_+=0.90$ to $2.00.$ }
\label{axial_z2}
\end{figure}
\begin{figure}[!t]
\begin{center}
\includegraphics[angle=0]{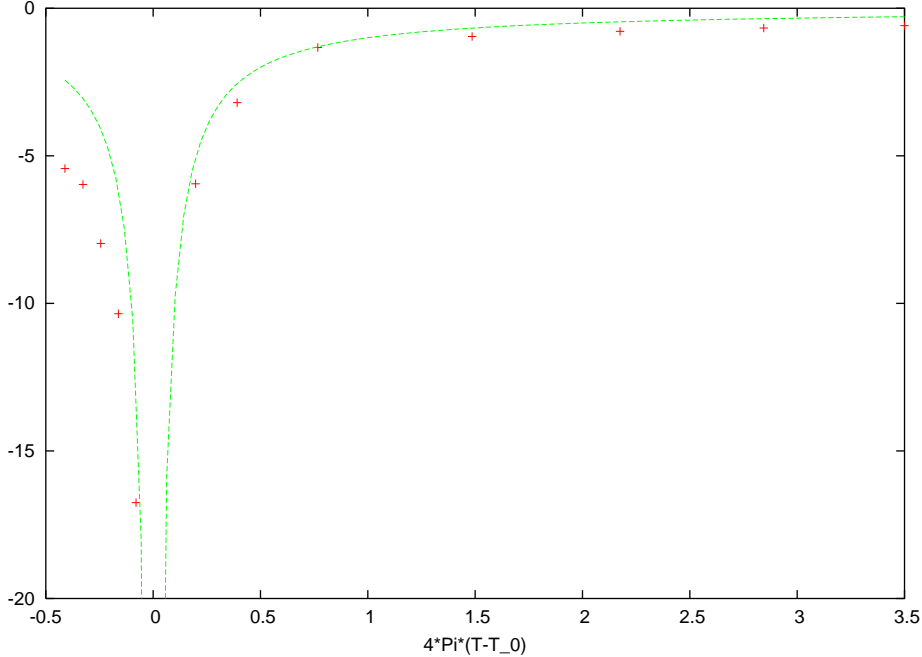}
\end{center}
\caption{The imaginary part of the lowest frequency $\omega_0$ of
the axial $Z_2$ purely dissipative mode versus temperature for
$\xi = 1$. Also shown is a fit $\Im\omega_0 = -\frac{1}{4\pi
|T-T_0|}$. \label{anafig4}}
\end{figure}
\begin{figure}[!t]
\centering
\includegraphics[angle=-90,scale=0.3]{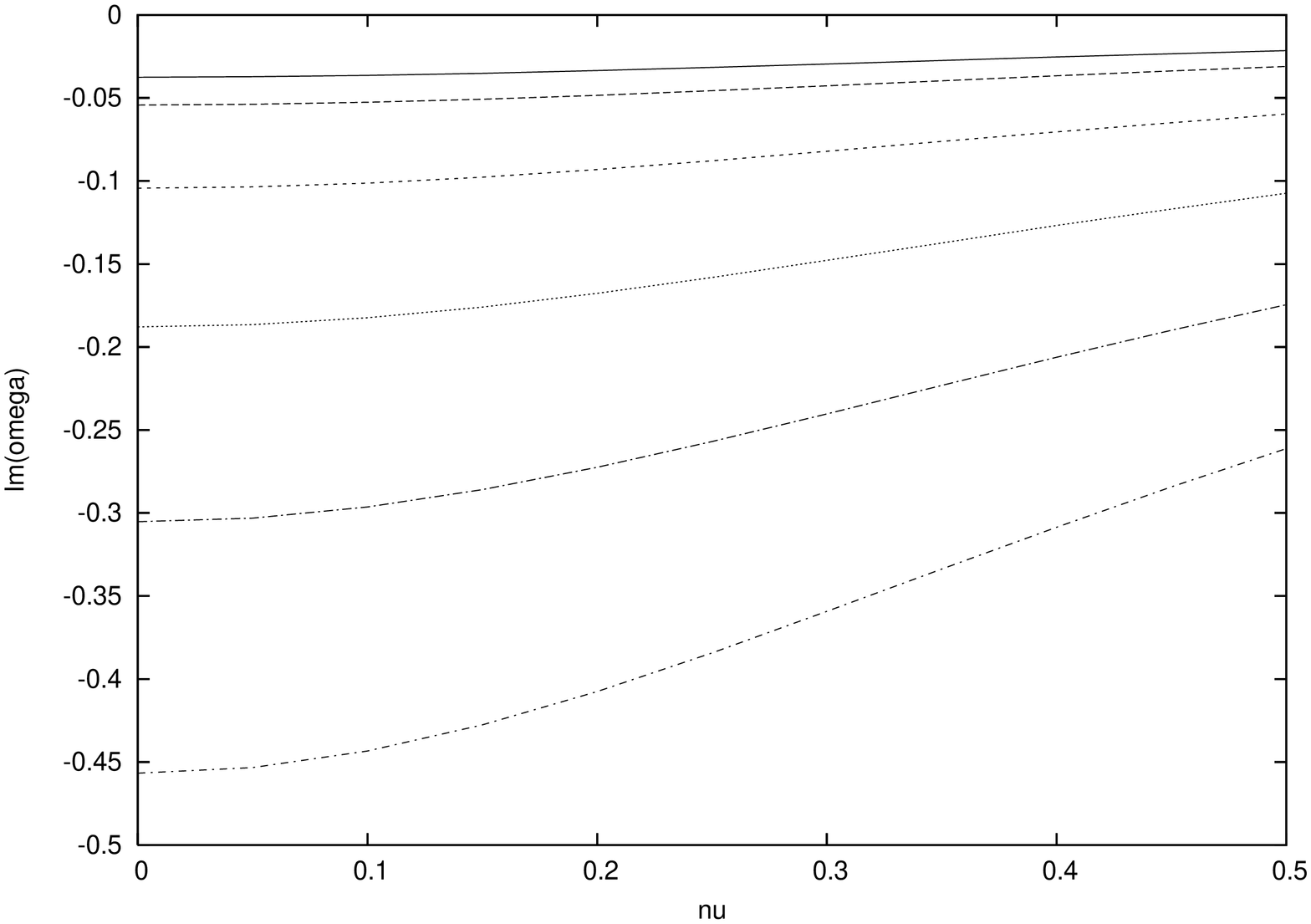}

\includegraphics[height=2in,width=4in]{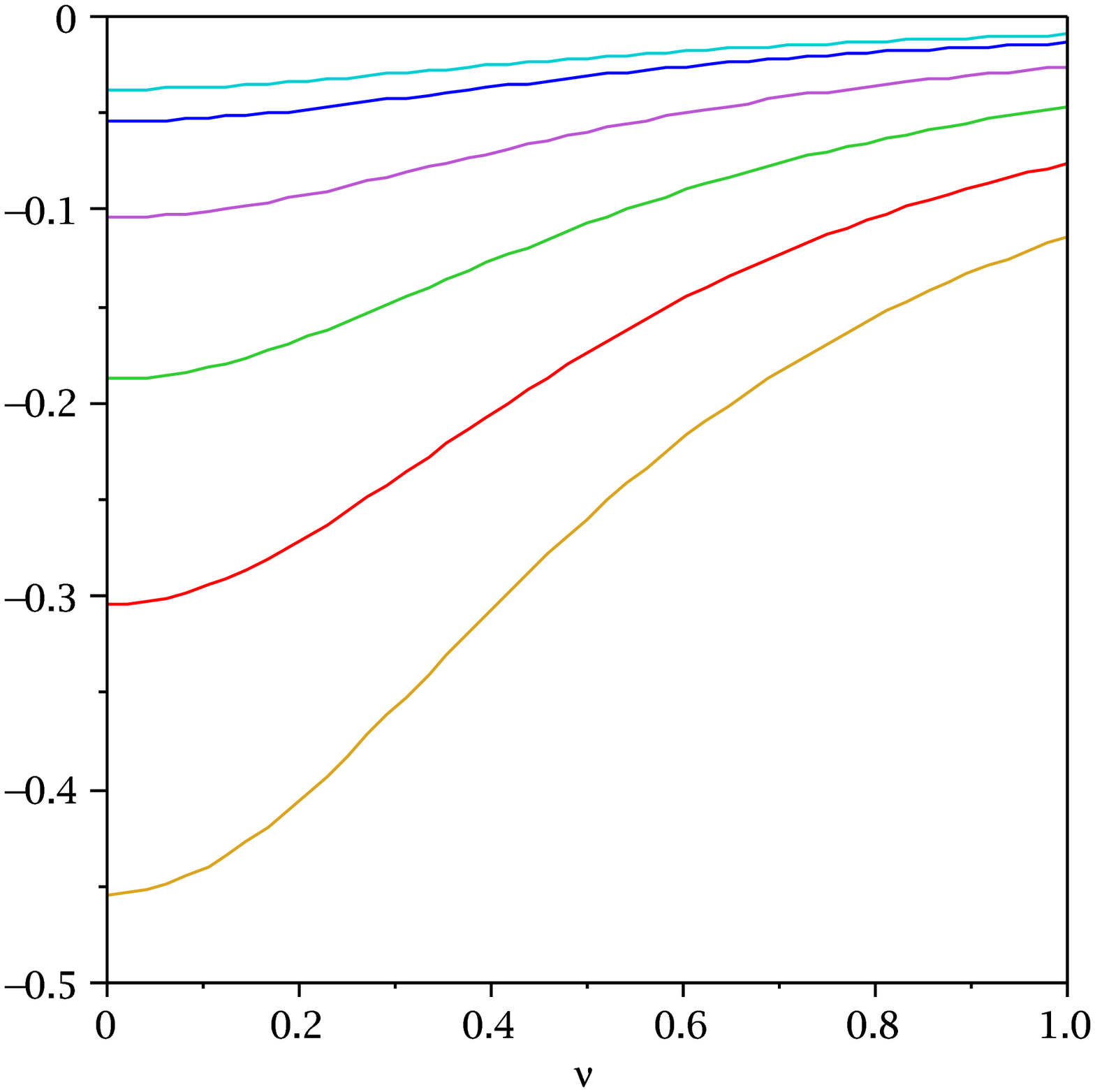}
\caption{$\Im\omega$ versus fractional charge $\nu$ for the lowest
purely dissipative mode of axial $Z_2$ perturbations at
$r_+=20.00$. The corresponding $\lambda$ is smaller than $0.15.$
The curves correspond to (top to bottom) $\xi=0,1 ,2, 3, 4, 5.$
The second panel shows the analytical expression for comparison.}
\label{tImo}
\end{figure}
\begin{figure}[!t]
\centering
\includegraphics[angle=-90,scale=0.3]{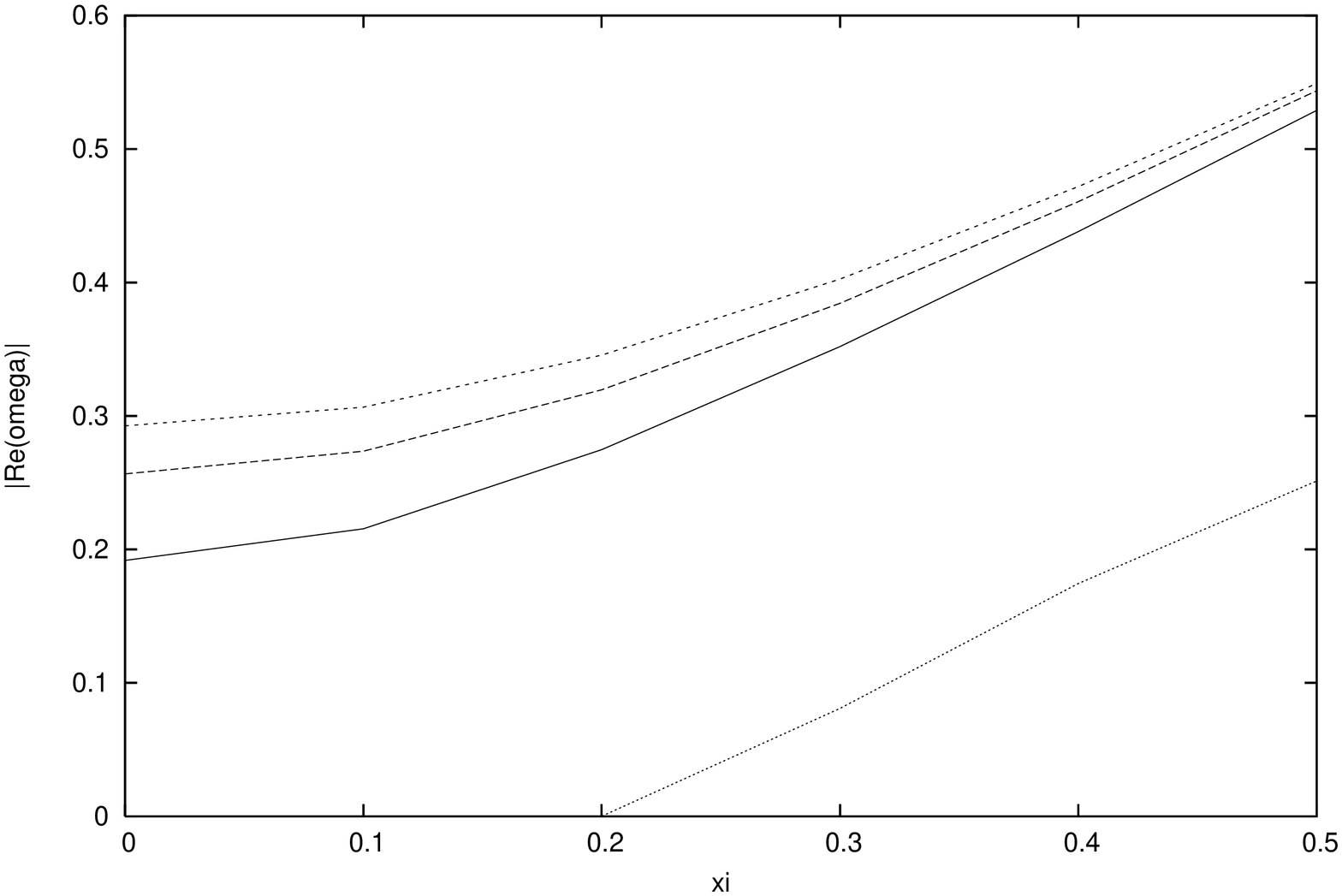}

\includegraphics[height=2in,width=3in]{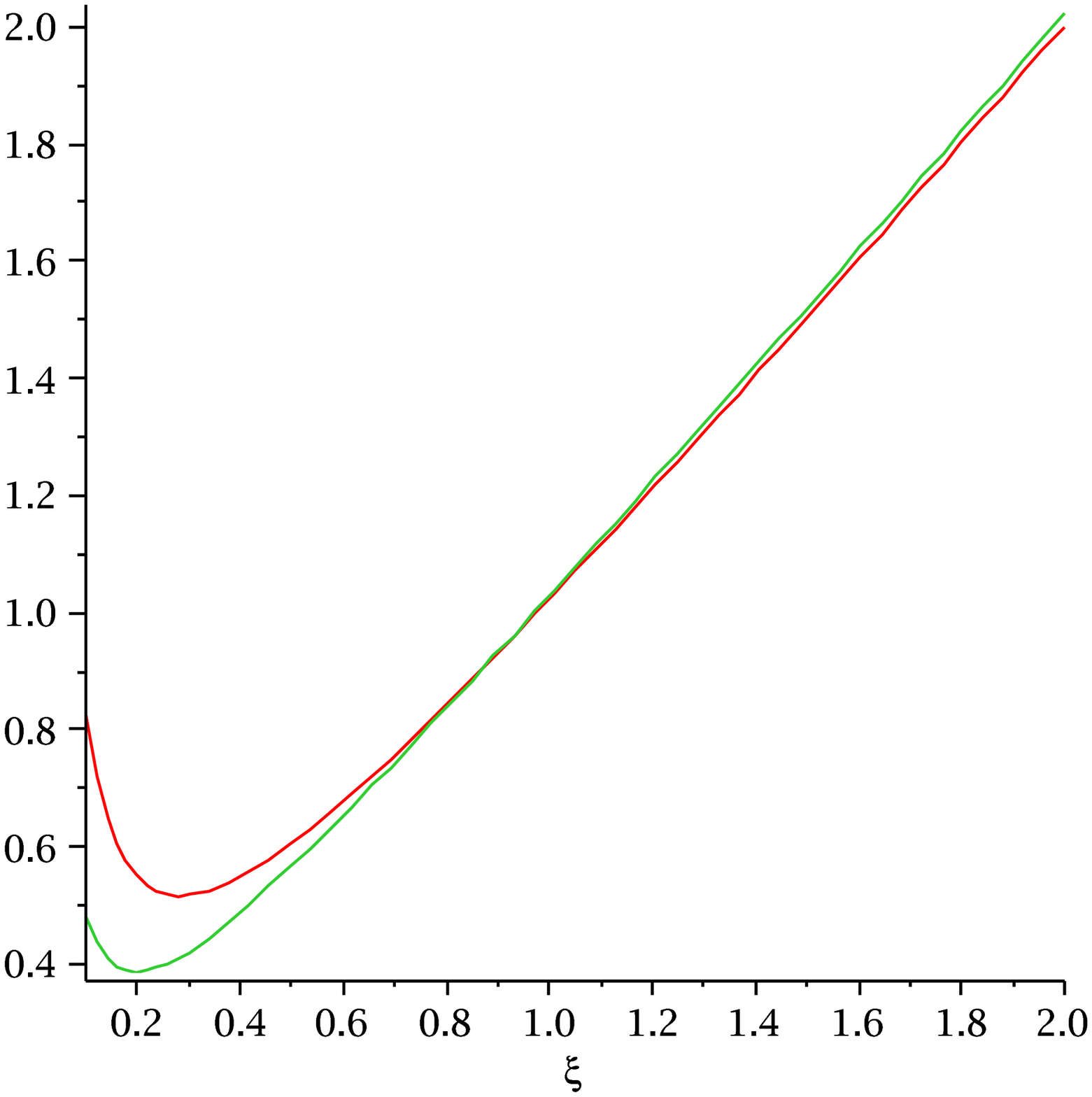}
\caption{The (absolute values of the) real parts of $Z_2$ modes
versus $\xi$ at $r_+=0.995.$ The curves correspond to the values
(top to bottom) $\Im\omega = -5.45, -3.47, -1.49, -29.21$. The
second panel shows the analytic approximation for the lowest two
modes.
} \label{z2_0995}
\end{figure}
\begin{figure}[!b]
\centering
\includegraphics[angle=-90,scale=0.3]{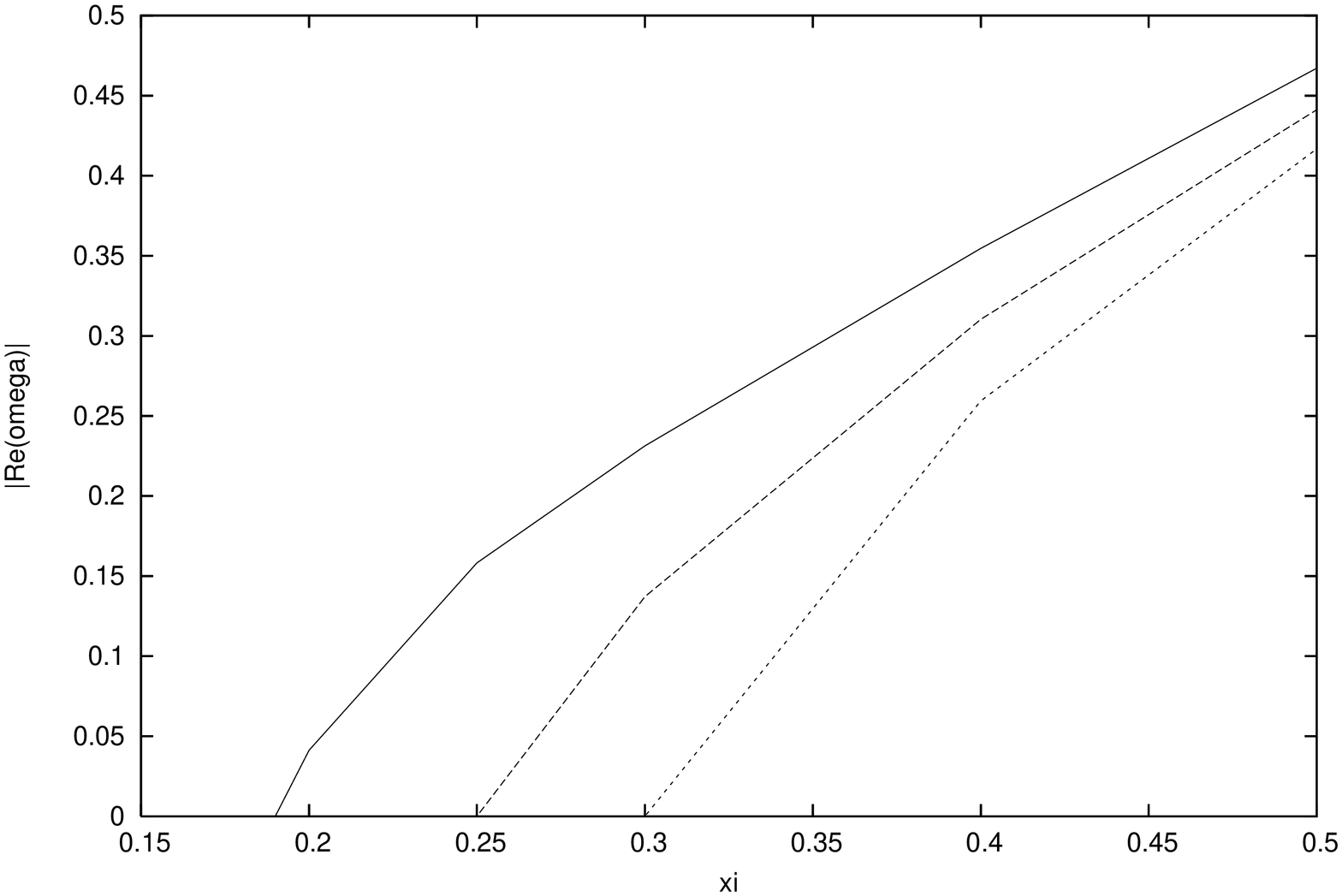}

\includegraphics[height=2in,width=3in]{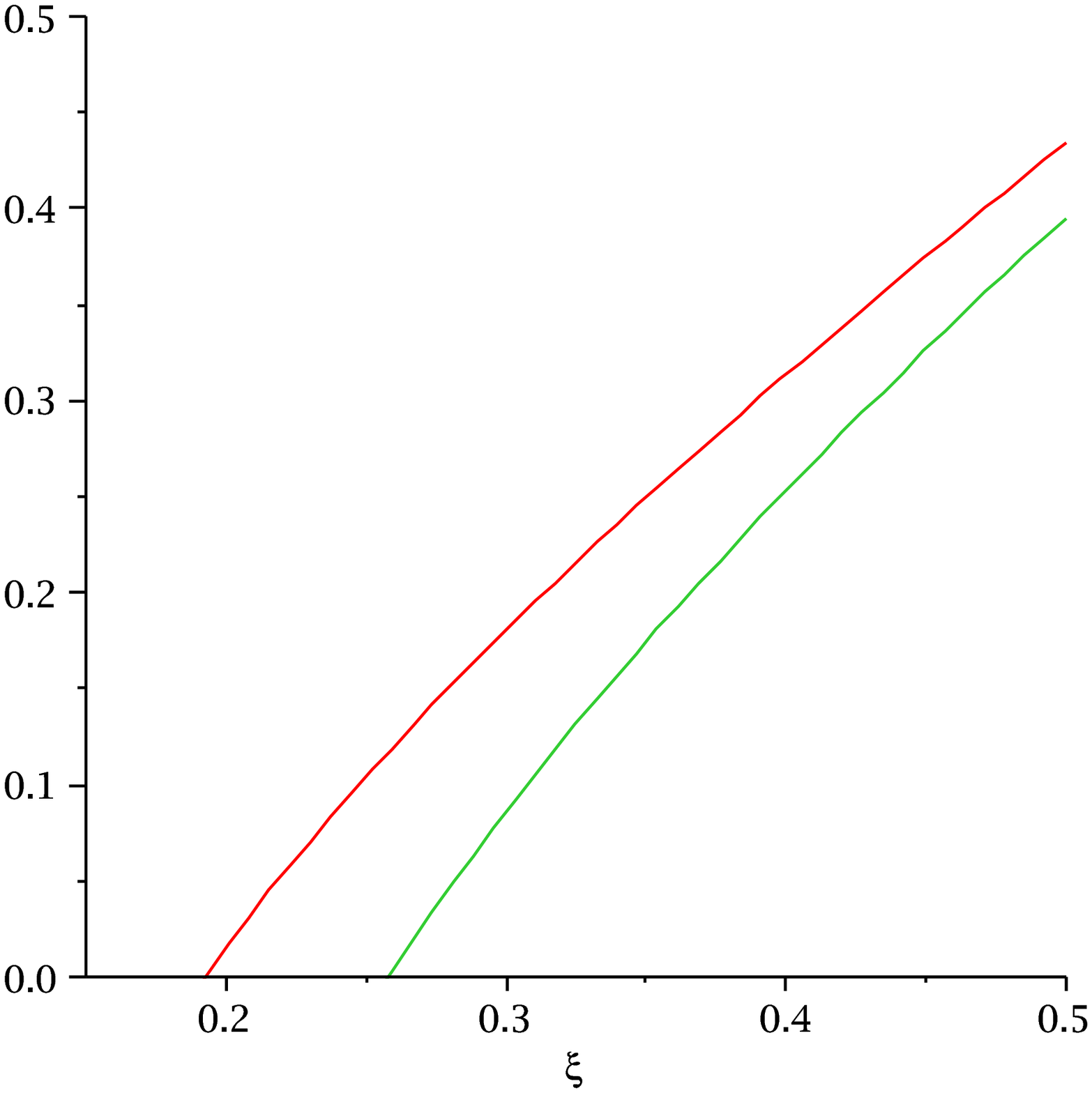}
\caption{The (absolute values of the) real parts of the $Z_2$
axial modes versus $\xi$ at $r_+=1.005$, $\nu =0$. The imaginary
part decreases from left to right. The second panel shows the
analytic estimate (eq.~(\ref{eqar1})) for the two lowest modes.}
\label{z2_1005}
\end{figure}
\begin{figure}[!t]
\begin{center}
\includegraphics[angle=0]{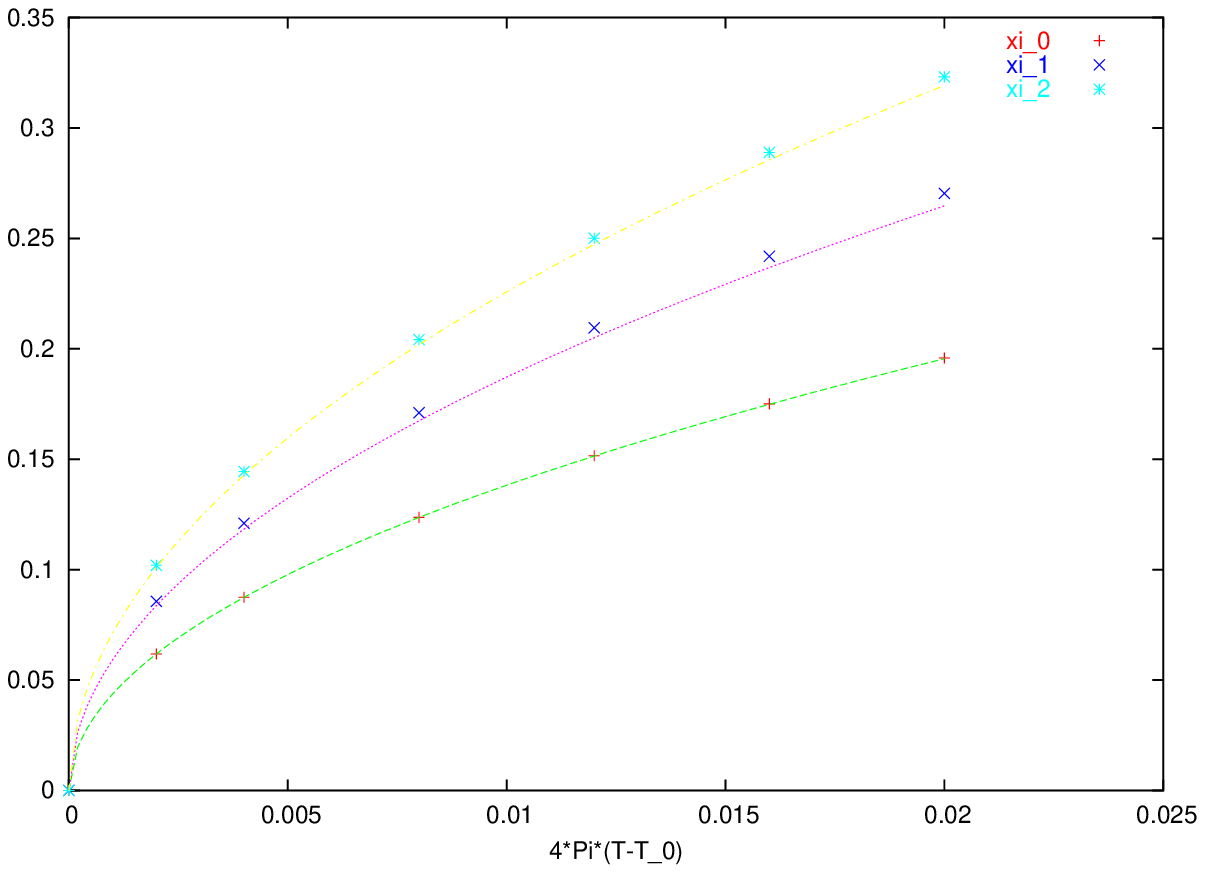}
\end{center}
\caption{The first three critical values of $\xi$ for axial $Z_2$
modes versus temperature for $\nu = 0$. Also shown are fits,
$\xi_0 = \sqrt{24 (T-T_0)}$, $\xi_1 = \sqrt{46 (T-T_0)}$, $\xi_2 =
\sqrt{64 (T-T_0)}$. \label{anafig2}}
\end{figure}

We observe that the real parts decrease with decreasing $\xi$
until they vanish at some critical value. For $\xi$ smaller than
this critical value, the QNMs will be pure imaginary. The critical
velues of $\xi$ given by the analytic expressions (\ref{eqana4})
are in good agreement with numerical results. This is seen in
Fig.~\ref{anafig1} which shows the singular behaviour of the three
lowest critical values of $\xi$ just below the critical point
$T=T_0$. The data points have been calculated both numerically and
by using the analytic equation (\ref{eq54}).

For a given $\xi$, the corresponding modes will be purely
dissipative  for large enough $\Im\omega,$ since the relevant
curve will have crossed the axis at a value greater than the given
$\xi$; however there will exist in general curves (that is,
appropriate values of $\Im\omega$), for which the corresponding
$\Re\omega$ will be different from zero and the QNM will be a
propagating one. Therefore, QNMs with a sufficiently large
absolute value of the imaginary part will be purely dissipative,
while the lowest QNMs will be propagating. This calculation
confirms once again that the lowest QNMs will be propagating,
followed by purely dissipative ones. This fully agrees with the
numerical results for small horizons and zero charge. The
situation is exactly the opposite for large horizons: in that case
the curves of approximately constant imaginary part never cross
the horizontal axis. For non-zero values of the charge the picture
is similar.

\subsection{Axial $Z_2$ Perturbations}

The results are described starting in Fig.~\ref{z2_150} which
provides a picture similar in some respects to the corresponding
result for $Z_1$ excitations shown in Fig.~\ref{150_500}. Two
differences should be noted, however: (a) There is no qualitative
distinction between intermediate and large horizons here. In
particular, there is no need to also depict the behaviour of
$r_+=5.00,$ as we did in Fig.~\ref{150_500}, and (b) the absolute
value of the lowest QNM is much smaller than the remaining ones;
it almost coincides with the horizontal axis.

For small horizons we have chosen to depict the results for
$r_+=0.995$ in Fig.~\ref{0995}. The picture is qualitatively the
same as the corresponding result for $Z_1$ excitations, shown in
Fig.~\ref{z2_995}. In particular, nothing is special about the
lowest mode in this case: this should be contrasted with the
results above the critical point presented previously in
Fig.~\ref{z2_150}.

\subsubsection{Temperature Dependence}

Next we examine the temperature (or horizon) dependence of axial
$Z_2$ modes. Fig.~\ref{axial_z2} contains the numerical results
for the two lowest purely dissipative modes versus $T-T_0$ for
$\lambda=0.50.$ The lowest mode has been reported
in~\cite{CL,kokkotas} to scale as $\fr{1}{r_+}$ which agrees with
our analytic asymptotic expression (\ref{eqZ2ana}). Using the
temperature, rather than the horizon, as an independent variable,
we have found that this mode scales as $a+\fr{b}{T-T_0}$ above the
critical temperature; this fit is of very good quality and is
shown in the figure. However, we have not been able to fit the
$T<T_0$ data to a function of this form except very close to the
critical point $T=T_0$. There is an infinite change in slope at
the critical temperature, as in the $Z_1$ case. We also remark
that for $T<T_0$ the two lowest values are very close to each
other, while for $T>T_0$ we have a lowest mode with a very small
absolute value, while the absolute value of the second lowest mode
is much larger.

The above conclusions are confirmed by our analytic results. In
Fig.~\ref{anafig4}, we show the lowest mode for $\lambda = 0$ for
various temperatures both above and below the critical point. The
data points were found by solving the analytic equation
(\ref{eq57}). Once again, we observe a singular behaviour
characterized by an infinite change in the slope at the critical
point.

\subsubsection{$\xi-$Dependence}

We now proceed with a discussion of the $\xi-$dependence for the
two regimes: large and small horizons. Fig.~\ref{tImo} contains
both numerical and analytic results (using the asymptotic
expression (\ref{eqZ2ana})) for $r_+=20.00$ The value $\xi=0$
yields the uppermost curve. This behaviour is similar to the
behaviour for intermediate horizons in the axial $Z_1$ case
(Fig.~\ref{intermh_xi_q=040}) and is quite different from the
behaviour for large horizons in that case
(Fig.~\ref{largeh_xi_q=0}).

For horizons below the critical point, results are contained in
Fig.~\ref{z2_0995} where we chose a value close to the critical
point, $r_+ = 0.995$. The results are slightly different from the
$Z_1$ case (Fig.~\ref{z2xiplt}). In fact, the real part initially
increases, it attains a maximal value and then it decreases and
cuts the axis. This happens for very large values of the imaginary
part, of the order of 30, the exact value depending on $\xi.$
Coming back to Fig.~\ref{z2_0995}, the lowest plotted curve
corresponds to such a value. We have also plotted the analytic
results (\ref{eqana1}) and (\ref{eqar1}). The agreement with the
numerical results is good when $\xi$ is not small. As $\xi\to 0$,
the corrections (\ref{eqar1}) exceed the zeroth order result
(\ref{eqana1}) and the first-order approximation fails.

For comparison, we also show results slightly above the critical
point in Fig.~\ref{z2_1005} where we chose $r_+ = 1.005$ and $\nu
=0$. Here the behaviour is very similar to the behaviour of $Z_1$
modes slightly {\em below} the critical points ({\em cf.}~with
Fig.~\ref{z2xiplt}). Similar remarks can be made in this case. For
each mode, there exists a critical value of $\xi$ below which the
mode does not propagate and becomes purely dissipative ($\Re\omega
= 0$). These critical values of $\xi$ depend on the temperature.
Their behaviour near the critical tempetarure is given by the
analytic expressions (\ref{eqana6a}), (\ref{eqana6b}) and plotted
in Fig.~\ref{anafig2} ({\em cf}~with Fig.~\ref{anafig1} for $Z_1$
modes). Data points were found numerically as well as by solving
the analytical equation (\ref{eq57}).

\section{Conclusions}
\label{sec6}

We have studied the perturbative behaviour of the charged
topological black holes. We have calculated both analytically and
numerically the QNMs of electromagnetic and gravitational
perturbations of these black holes.

For large black holes we found analytically that the axial $Z_{2}$
QNMs are purely dissipative, depend on the charge and scale as the
inverse of the black hole horizon. The $Z_{1}$ axial modes are
proportional to the black hole horizon but analytical expressions
cannot be obtained in general for non-zero charge. For zero charge
the potentials for both axial and polar modes reduce to
electromagnetic potential and the wave function can be written in
terms of the Heun function, leading to a semi-analytic expressions
of the QNMs.

For small black holes, at the critical point with zero charge and
mass the wave equation simplifies and the QNMs can be explicitly
calculated. For small changes around the critical point, the real
part of the $Z_{2}$ modes increases above the critical point
giving a positive slope, whereas below the critical point it gives
a negative slope.  Above the critical point, for these modes there
is a critical value of $\xi$ below which there are only purely
dissipative modes. Below the critical point, there are no purely
dissipative modes for any value of $\xi$. The $Z_{1}$ modes
exhibit the opposite behaviour. Similar behaviour is exhibited by
the polar modes.

These results are also supported by  numerical investigations of
the QNMs. The numerical results show clearly a change of slope of
QNMs around a critical temperature for all kinds of perturbations.
We found that the purely dissipative modes scale linearly with
temperature for large black holes, while for small horizons they
scale according to $a+b/(T-T_{0})$. Then, for a fixed charge to
mass ratio we observed an infinite change of slope at $T=T_{0}$
signaling a second order phase transition.

The numerical results show also an interesting dependence of the
modes on the charge of the black hole. For small horizons and
small charge the number of propagating ($\Re\omega \ne 0$) QNMs is
finite, while as the charge increases, positive slope frequencies
coexist with frequencies of negative slope and the number of
propagating QNMs is again infinite. As the charge increases, a
drastic change in the temperature dependence occurs.

\newpage

\section*{Acknowledgments}

Work supported by the NTUA research program PEVE07. E.~P.~was
partially supported by the European Union through the Marie Curie
Research and Training Network UniverseNet (MRTN-CT-2006-035863).
G.~S.~was supported in part by the US Department of Energy under
grant DE-FG05-91ER40627.

\end{document}